\documentclass[lettersize,journal]{article}
\usepackage{amsmath,amsfonts}
\usepackage{algorithmic}
\usepackage{algorithm}
\usepackage{array}
\usepackage[caption=false,font=normalsize,labelfont=sf,textfont=sf]{subfig}
\usepackage{textcomp}
\usepackage{stfloats}
\usepackage{url}
\usepackage{verbatim}
\usepackage{cite}
\usepackage{amsthm,amsmath,amssymb}
\usepackage{mathrsfs}
\usepackage{graphicx}
\usepackage{pdfpages}
\usepackage{subfig}
\usepackage{geometry}
\usepackage{indentfirst}
\usepackage{multirow}
\usepackage{diagbox}
\usepackage{makecell}
\usepackage[section]{placeins}
\usepackage{authblk}
\usepackage{booktabs}
\usepackage{appendix}

\usepackage[export]{adjustbox}
\begin{document}
	
	\title{Network Structure Governs Drosophila Brain Functionality}
	\author[1]{Xiaoyu Zhang}
	
	\author[1]{Pengcheng Yang}
	\author[1]{Jiawei Feng}
	\author{Kang Wen}
	\author[2]{Qiang Luo}
	\author[2]{Wei Lin}
	\author[1]{Xin Lu\thanks{Corresponding author: xin.lu.lab@outlook.com}}
	
	\affil[1]{College of Systems Engineering, National University of Defense Technology, Changsha 410073, China}
	\affil[2]{Research Institute of Intelligent Complex Systems, Fudan University, Shanghai 200433, China}
	


\date{}
\maketitle

\begin{abstract}

	How intelligence emerges from living beings has been a fundamental question in neuroscience. However, it remains largely unanswered due to the complex neuronal dynamics and intricate connections between neurons in real neural systems. To address this challenge, we leveraged the largest available adult Drosophila connectome data set, and constructed a comprehensive computational framework based on simplified neuronal activation mechanisms to simulate the observed activation behavior within the connectome. The results revealed that even with rudimentary neuronal activation mechanisms, models grounded in real neural network structures can generate activation patterns strikingly similar to those observed in the actual brain. A significant discovery was the consistency of activation patterns across various neuronal dynamic models. This consistency, achieved with the same network structure, underscores the pivotal role of network topology in neural information processing. These results challenge the prevailing view that solely relies on neuron count or complex individual neuron dynamics. Further analysis demonstrated a near-complete separation of the visual and olfactory systems at the network level. Moreover, we found that the network distance, rather than spatial distance, is the primary determinant of activation patterns. Additionally, our experiments revealed that a reconnect rate of at least 1‰ was sufficient to disrupt the previously observed activation patterns. We also observed synergistic effects between the brain hemispheres: Even with unilateral input stimuli, visual-related neurons in both hemispheres were activated, highlighting the importance of interhemispheric communication. These findings emphasize the crucial role of network structure in neural activation and offer novel insights into the fundamental principles governing brain functionality.
	
	
\end{abstract}


\section{Introduction}

The neural systems of intelligent beings maintain continuously dynamical communication\cite{lerner2016communication}. 
This process gives rise to macroscopic emergence across the entire neural system. Such communication fosters the emergence of intelligence in all intelligent life forms. Consequently, comprehending and replicating this process has become an indispensable avenue for unraveling biological intelligence and realizing artificial intelligence\cite{zador2023catalyzing}.

Neural systems are often modeled as interconnected networks\cite{avena2018communication}, with neurons engaging in constant communication via synapses within the network's topology. To elucidate the relationship between the network itself and its dynamics in a neural network, various models have been developed, spanning from the cellular level to the whole brain. For example, neuronal dynamic models like Leaky Integrate-and-Fire (LIF) model\cite{lansky2008review} are designed to describe the electrical activity and dynamic behavior of neurons\cite{hodgkin1952quantitative}. These models are characterized by the following features: they typically use a set of differential equations to describe the electrophysiological behavior of neurons and usually contain a set of parameters that need to be measured experimentally or estimated.
On a larger scale, a vast array of isolated neuronal dynamics, ranging from several to millions, are interconnected and parameter-optimized to simulate the local or global dynamics of the brain\cite{zeng2023braincog}. Due to the requirements of commonly used algorithms like backpropagation on network structures, these methods typically do not fully adhere to the structures of real neural networks. Another category of network-based methods applies propagation models on complex networks to neural networks, thereby describing information propagation and emergent mechanisms within neural networks. These methods are collectively referred to as brain network communication models. Such approaches typically involve substantial simplification of neuron models, resulting in fewer parameters, but also leading to the loss of some electrophysiological details of neurons\cite{seguin2023brain}. These two seemingly opposing approaches illustrate that it remains unresolved whether the activation mechanisms of neurons or the connectivity between neurons play the more dominant role. This question is fundamental to understanding how intelligence emerges from neural systems. 

Additionally, in recent years, as an innovative way to modify intelligence, artificial neural networks (ANNs) have achieved significant advancements in numerous fields. \cite{greener2022guide}. Despite these achievements, the structure of ANNs does not strictly mimic the actual neural network structure of biological brains. This discrepancy raises questions about the hypothesis that the inherent structure of a network is crucial for intelligence. Furthermore, the scaling law in the field of machine learning posits that the more parameters an ANN has, the greater the degree of emergent intelligence it can exhibit\cite{kaplan2020scaling}. These theory hints that the quantity of neurons (parameters) might be a more significant factor in the development of intelligence than the precise arrangement of these neurons in a network, suggesting that the specific network structures found in biological systems may not be essential for the emergence of intelligence.
The observation of neural network structures is crucial for interpreting intelligence from a network perspective. Current neuroimaging techniques, such as Magnetic Resonance Imaging (MRI) and functional MRI (fMRI), offer non-invasive exploration but lack the spatial resolution necessary to resolve individual neurons and synapses. However, these measurement methods face challenges in obtaining information about the positions and connectivity of neurons in the nervous system, making it difficult to understand the nervous system at the neuron-synapse level\cite{logothetis2008we}. 
Employing electron microscopy (EM) and deep learning techniques for automated neuron reconstruction, brain atlases of a range of animals, including Drosophila larvae\cite{winding2023connectome}, have been mapped. A recent study\cite{dorkenwald2023neuronal} published an adult Drosophila connectome data set, comprising nearly 130,000 neurons and 50,000,000 synapses, with details of neuronal network structure at unforeseen resolution. With full proofreading of the central brain and approximately 80\% completion of the optic lobesis presented, the data represents the largest whole-brain connectome of an adult Drosophila to date and provides an unprecedented opportunity for comprehensive network analysis and topological characterization of the Drosophila brain.
Building on the outlined challenges and opportunities, this study utilized the aforementioned Drosophila connectome data set\cite{dorkenwald2023neuronal} to examine the importance of neuronal network's structure in simulating brain activation patterns. We constructed a large-scale network communication model framework based on simple mechanisms to simulate activation behavior observed in the connectome. Initially, we established a network representation of the connectome. We then developed a large-scale network communication model where neurons activate upon receiving sufficient stimuli and transmit signals to other neurons. This iterative process generated a global activation pattern. Our analysis employed threshold and sigmoid models as baselines, with a LIF model for comparison. We proposed an evaluation criterion based on the ratio of actual to expected neuron activations.

Additionally, we developed a 3D network visualization tool (see Appendix \ref{Visualization}), using HTML5's Canvas and Three.js based on WebGL technology. This software enables real-time observation of each neuron's communication process in three dimensional space. 








\section{Results}

\subsection{Structural Properties of Drosophila Connectome}
\begin{figure*}[!tbp]
		\centering
		\subfloat[]{
		\begin{minipage}[title]{\textwidth}
			\centering
			\includegraphics[width=0.9\textwidth, height=0.45\textwidth]{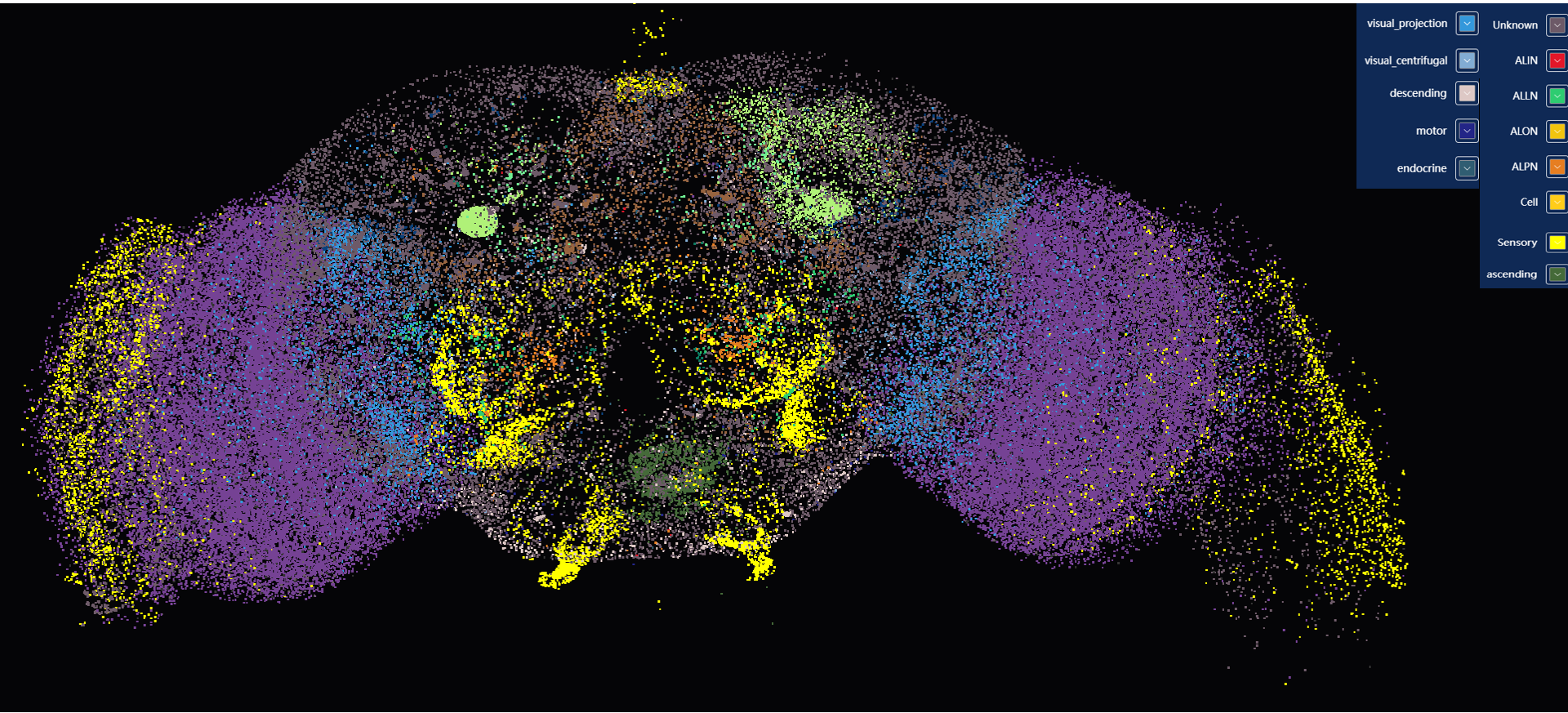}

		\end{minipage}
	}	
		
		\subfloat[]{
		\begin{minipage}[b]{0.3\textwidth}
			\centering
			\includegraphics[width=\textwidth,height=3.5cm]{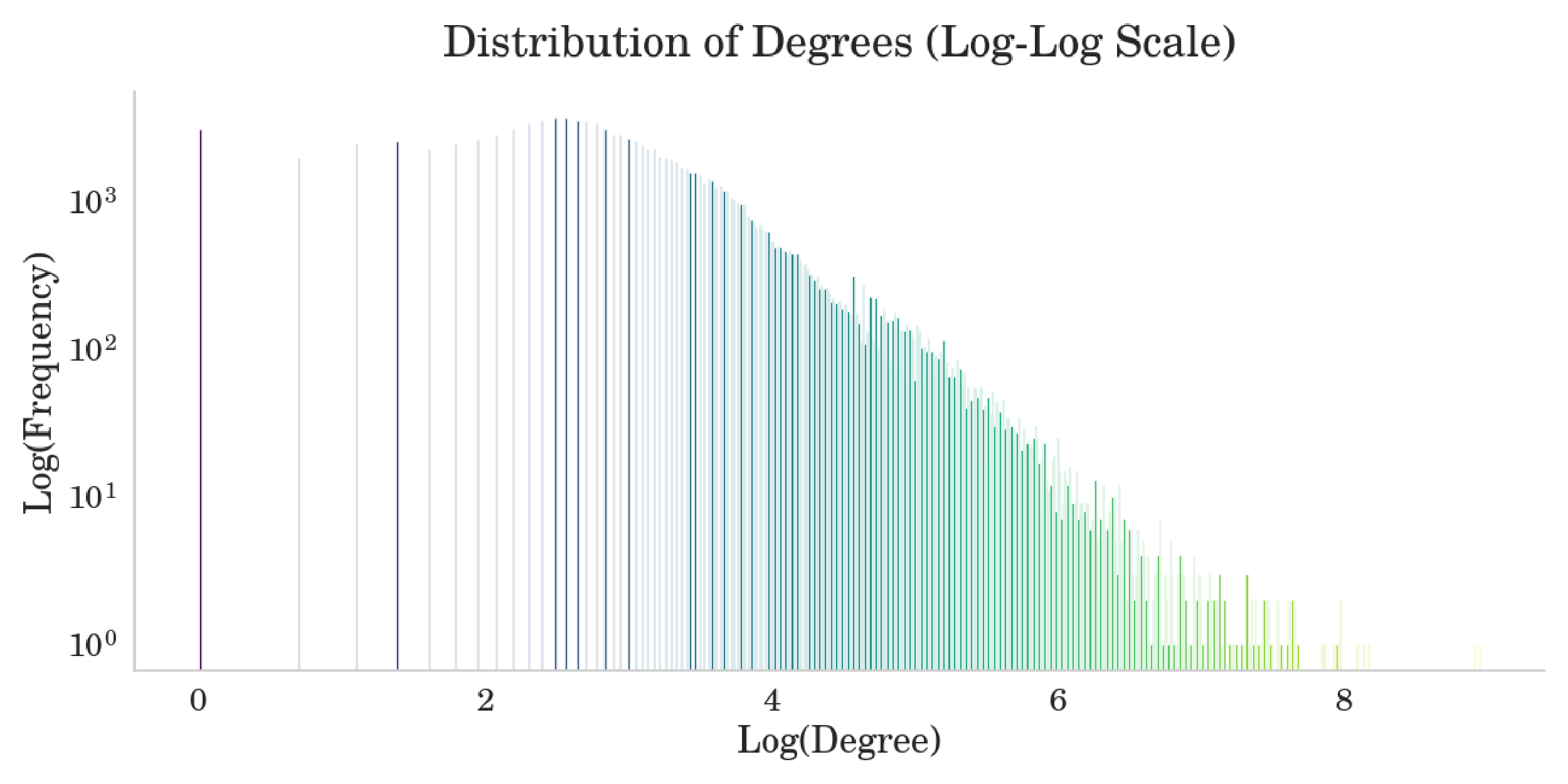}
			
		\end{minipage}
	}
	\subfloat[]{
		\begin{minipage}[b]{0.2\textwidth}
			\centering
			\includegraphics[width=\textwidth]{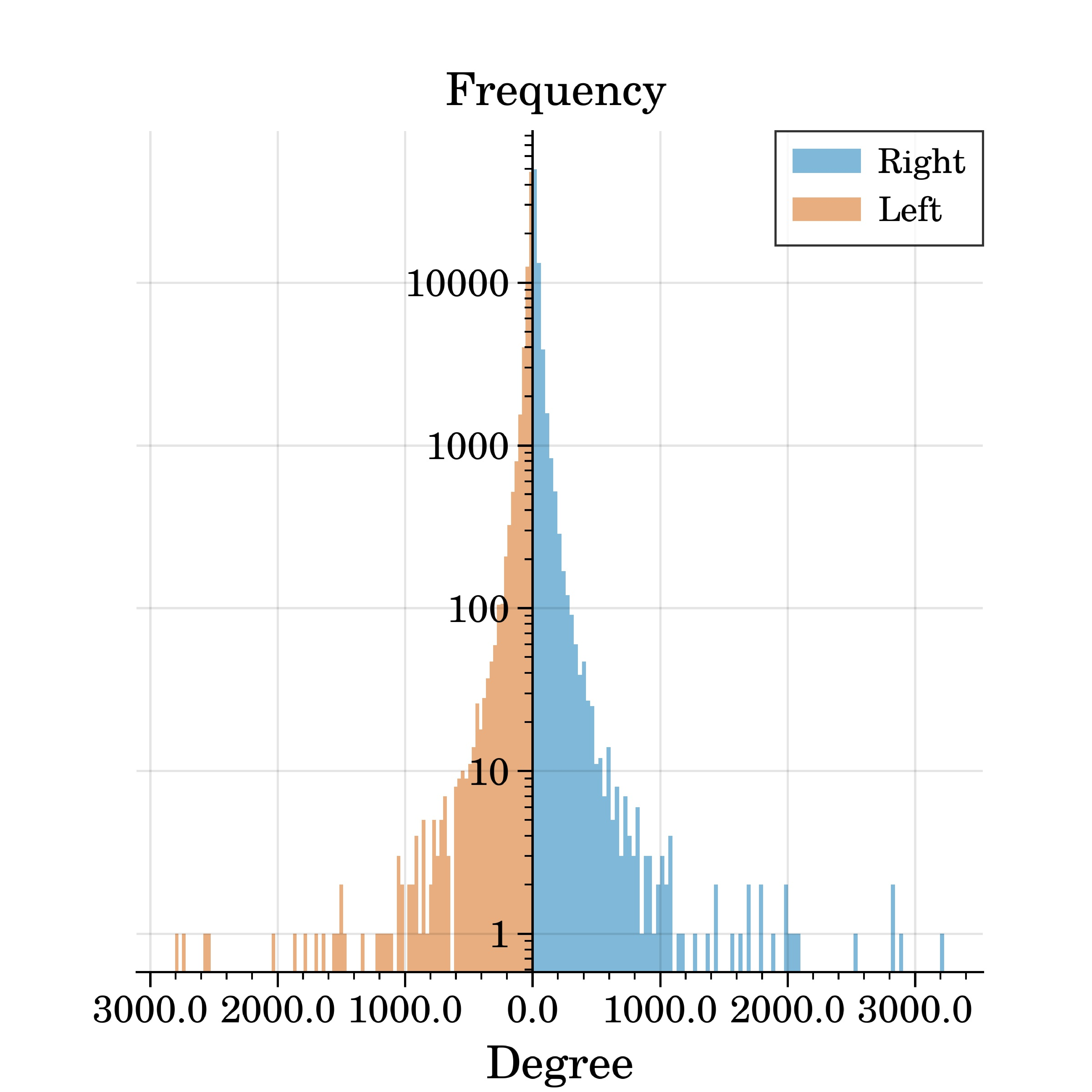}

		\end{minipage}
	}
	\subfloat[]{
		\begin{minipage}[b]{0.2\textwidth}
			\centering
			\includegraphics[width=\textwidth]{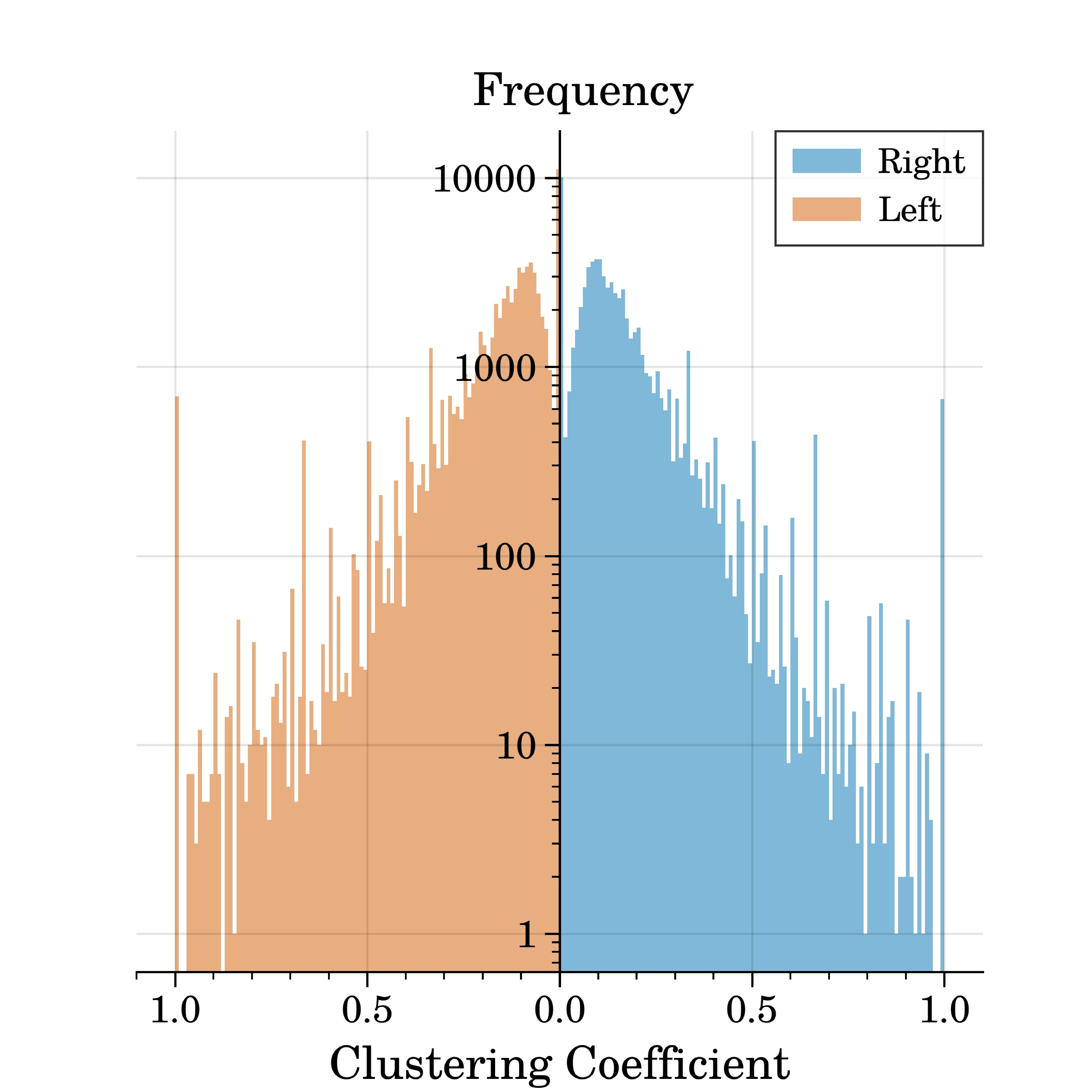}

		\end{minipage}
	}
	\subfloat[]{
		\begin{minipage}[b]{0.2\textwidth}
			\centering
			\includegraphics[width=\textwidth]{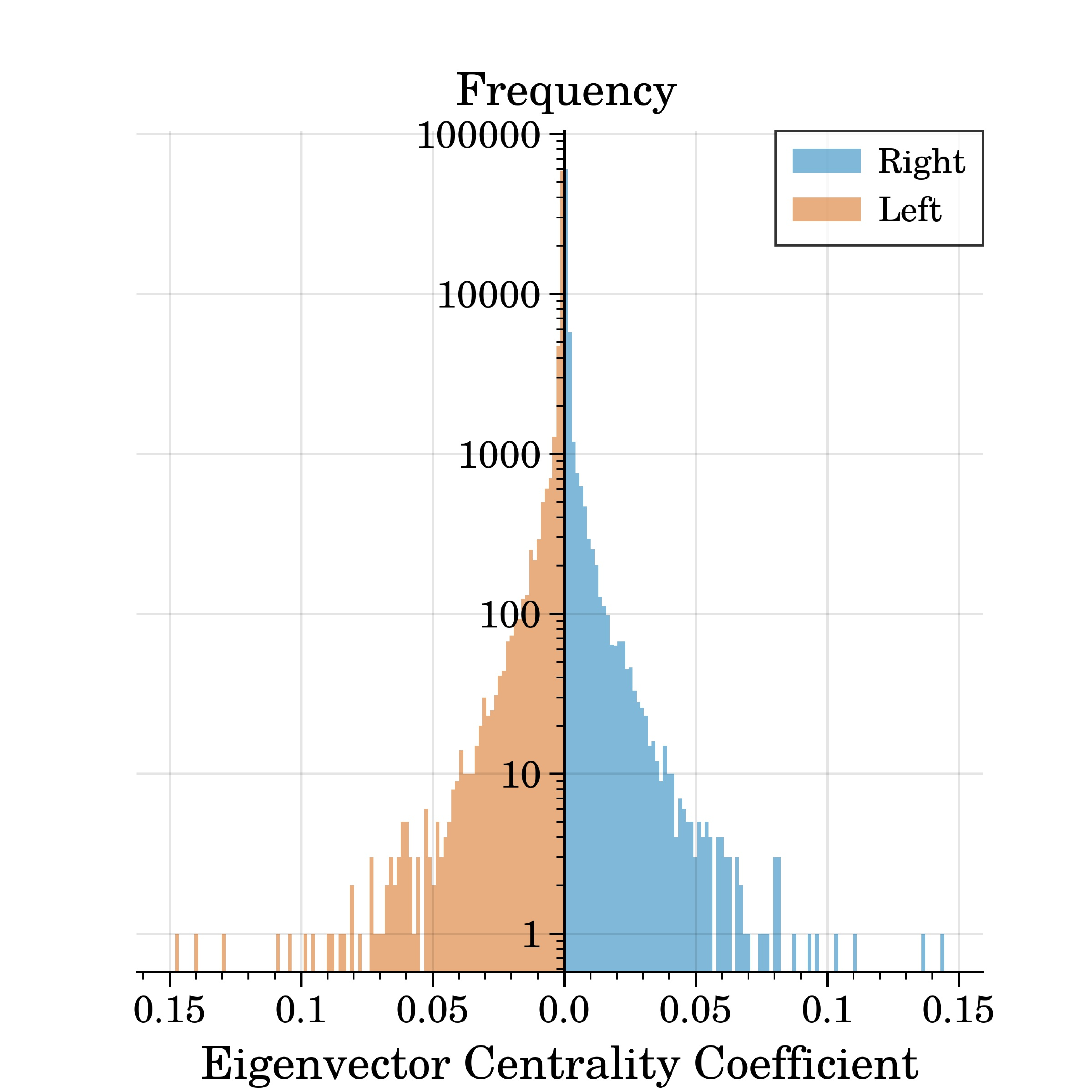}

		\end{minipage}
	}
	
		\subfloat[]{
		\begin{minipage}[b]{\textwidth}
			\centering
			\includegraphics[width=\textwidth]{./figures/cor_new.png}
			
	\end{minipage}}
	
	
		\caption{Visualization and statistic properties of Drosophila connectome data set. (a) is a visualization of Drosophila connectome. The different colors represent different functional regions. (b) shows the degree distribution of the network. (c), (d) and (e) plot the distribution of degree, clustering coefficient, and eigenvector centrality in each hemisphere, respectively. (f) represents the correlation between degree distribution and neuron characteristics. We use nodes to represent neurons' cell bodies and use edges to represent the synapses when calculated statistic properties. }\label{fig1}
\end{figure*}

The structure of the Drosophila connectome is presented in Fig. \ref{fig1}(a).  Visually, the shape of most of the functional regions of the brain appears to be physically similar. 
From Fig. \ref{fig1}(b) we can see that the degree, i.e., the number of connections to the nodes in the network, exhibits a long-tail distribution, implying that while a large number of neurons possess few synapses, a small number of neurons contains a large number of synapses. For example, more than 20\% of the nodes have a degree greater than 50, while only 0.3\% of the nodes have a degree greater than 500. This is consistent with previous research on functional brain networks\cite{chialvo2004critical}. 

To investigate whether the two hemispheres' network structures are similar in network statistic properties, we conducted a comprehensive network analysis on the two hemispheres. We calculated the degree, clustering coefficient, and eigenvector centrality distribution for every node in each hemisphere. 
According to Fig. \ref{fig1}(c)-(e), these statistical indicators for both hemispheres are extremely similar, indicating that the brain exhibits strong symmetry in network structure.


Also, we plotted the correlation curves between degree and cell length, cell surface area, and cell size (Fig. \ref{fig1}(f)). The significance coefficients \( R \) are 0.85 (\( p = 0.00 \)), 0.86 (\( p = 0.00 \)), 0.80 (\( p = 0.00 \)), respectively. The results indicate that the degree is logarithmically linearly correlated with these characteristics. This logarithmic relationship reveals a fundamental principle in the organization of neural networks in the brain.

\renewcommand\floatpagefraction{.99}
\renewcommand\topfraction{.99}
\renewcommand\bottomfraction{.99}
\renewcommand\textfraction{.1}
\setcounter{totalnumber}{50}
\setcounter{topnumber}{50}
\setcounter{bottomnumber}{50}

\subsection{Activation Patterns under Different Type of Stimuli}
In this section, we investigate the impact of different activation mechanisms on network activation patterns by utilizing various network activation mechanisms and observing the changes in activation rates across different areas of the network. Simultaneously, we examine the effect of global perturbations in network structure on activation patterns by altering the network architecture.
The optic and visual\_projection area are related to vision. ALLN, ALIN, ALON and ALPN are related to olfaction. Other supplemental messages of related neurons are presented in Appendix \ref{neuron}. 

Fig. \ref{scatter} illustrates the activation patterns of the large-scale network communication model (LSNC) using threshold, sigmoid, and LIF activation mechanisms. We also evaluated the model's performance after introducing perturbations. For a detailed description of these models, please refer to Appendix \ref{Model}. 

From left subgraphs of Fig. \ref{scatter} (a) and (b), it can be observed that the network model generates quasi-real activation patterns under specific stimuli: It is evident that visual (olfactory) stimuli result the most significant activation on visual (olfactory) neurons such as optic and visual projection neurons (ALLN, ALIN, ALPN, ALON neurons) while activating other neuron types minimally. 


Our findings indicate that different activation mechanisms can yield comparable activation patterns. In response to visual stimuli, all three models displayed notable activation in visual-related areas, whereas other areas exhibited minimal activation (left subgraphs of Fig. \ref{scatter} (a), (c), and (e)). The optic area, for instance, recorded activation rates of 36.39\%, 4.69\%, and 9.69\%. Under olfactory stimuli, the network also exhibits similar activation patterns in olfactory-related areas, with average activation rates of 94.07\%, 78.01\%, and 69.75\%, respectively (left subgraphs of Fig. \ref{scatter} (b), (d) and (f)). This suggests that the connectivity between neurons is more critical than the activation mechanisms of individual neurons. 

However, in the perturbed models (right subgraphs of Fig. \ref{scatter} (a)-(f)), the original quasi-real activation patterns disappear. No significant differences in activation regions were observed across all activation modes and types of stimuli. For instance, in the model with threshold activation mechanism, the maximum difference of all areas in the activation ratios under visual stimuli was only 3.162\%, and the range of regional activation ratios versus the average activation ratio was only 6.32\% (right subgraph of Fig. \ref{scatter}(a)), which indicates that the heterogeneity between different regions completely disappears. Additionally, in the sigmoid and LIF models, no significant activation was observed in any area.

Despite variations in activation rates among the three models, the substantial disparity between the activation ratios of relevant and irrelevant areas suggests the emergence of comparable quasi-real activation patterns across all models.

 \renewcommand{\dblfloatpagefraction}{.99}
\begin{figure*}[th]
	\centering  
	\vspace{-0.35cm} 
	
	\subfloat[threshold, visual]{
			\begin{minipage}[t]{0.5\linewidth}
					\includegraphics[width=7cm,height=3.5cm]{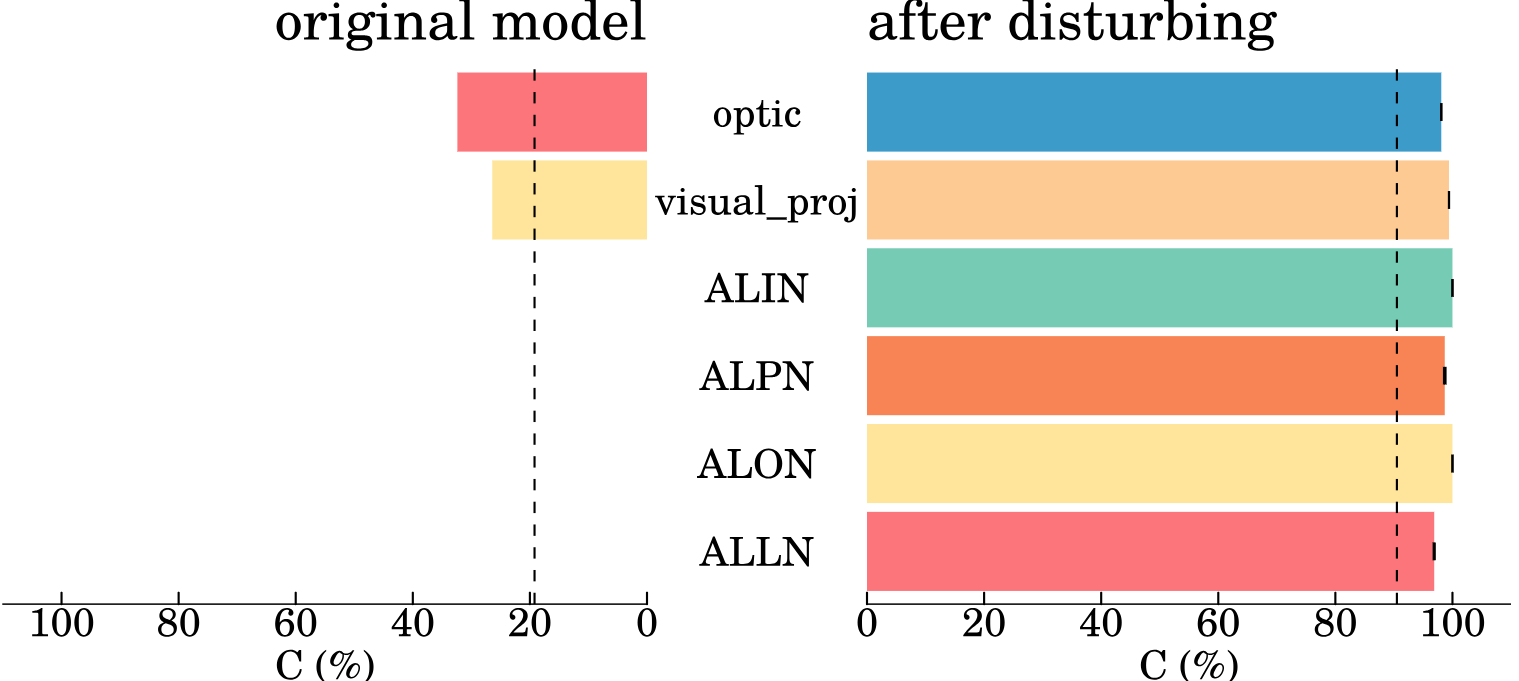}
				\end{minipage}		
		}
	\subfloat[threshold, olfactory]{
		\begin{minipage}[t]{0.5\linewidth}
				\includegraphics[width=7cm,height=3.5cm]{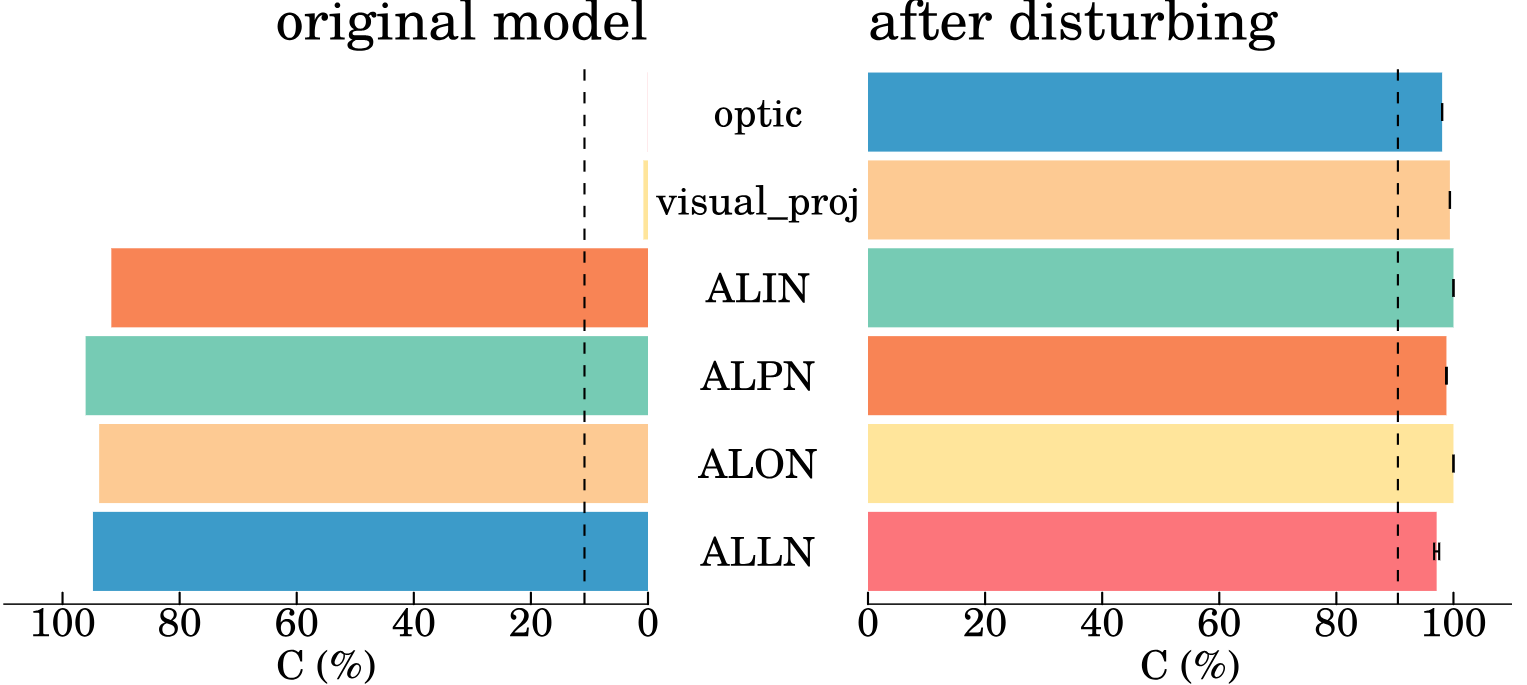}
			\end{minipage}		
		}
	\\
	\subfloat[sigmoid, visual]{
		\begin{minipage}[t]{0.5\linewidth}
				\includegraphics[width=7cm,height=3.5cm]{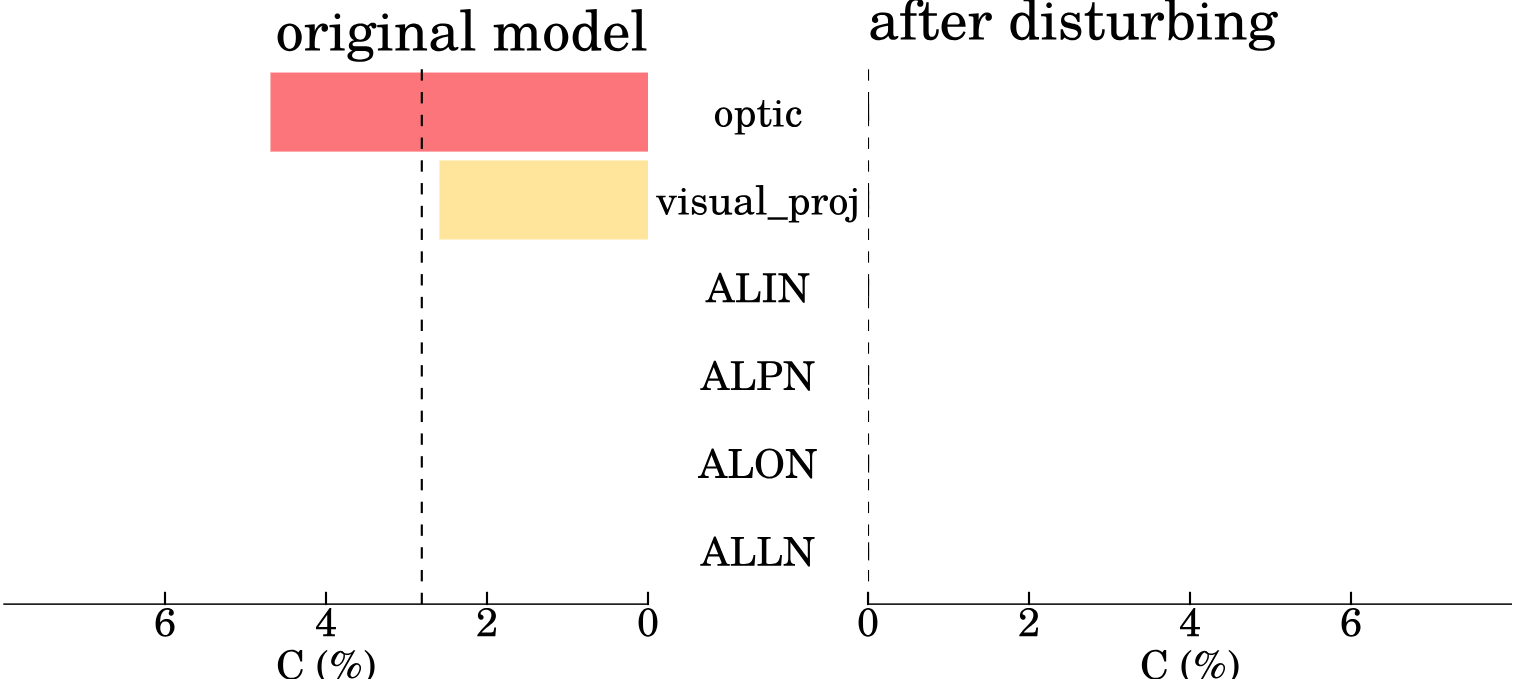}
			\end{minipage}		
		}
	\subfloat[sigmoid, olfactory]{
		\begin{minipage}[t]{0.5\linewidth}
				\includegraphics[width=7cm,height=3.5cm]{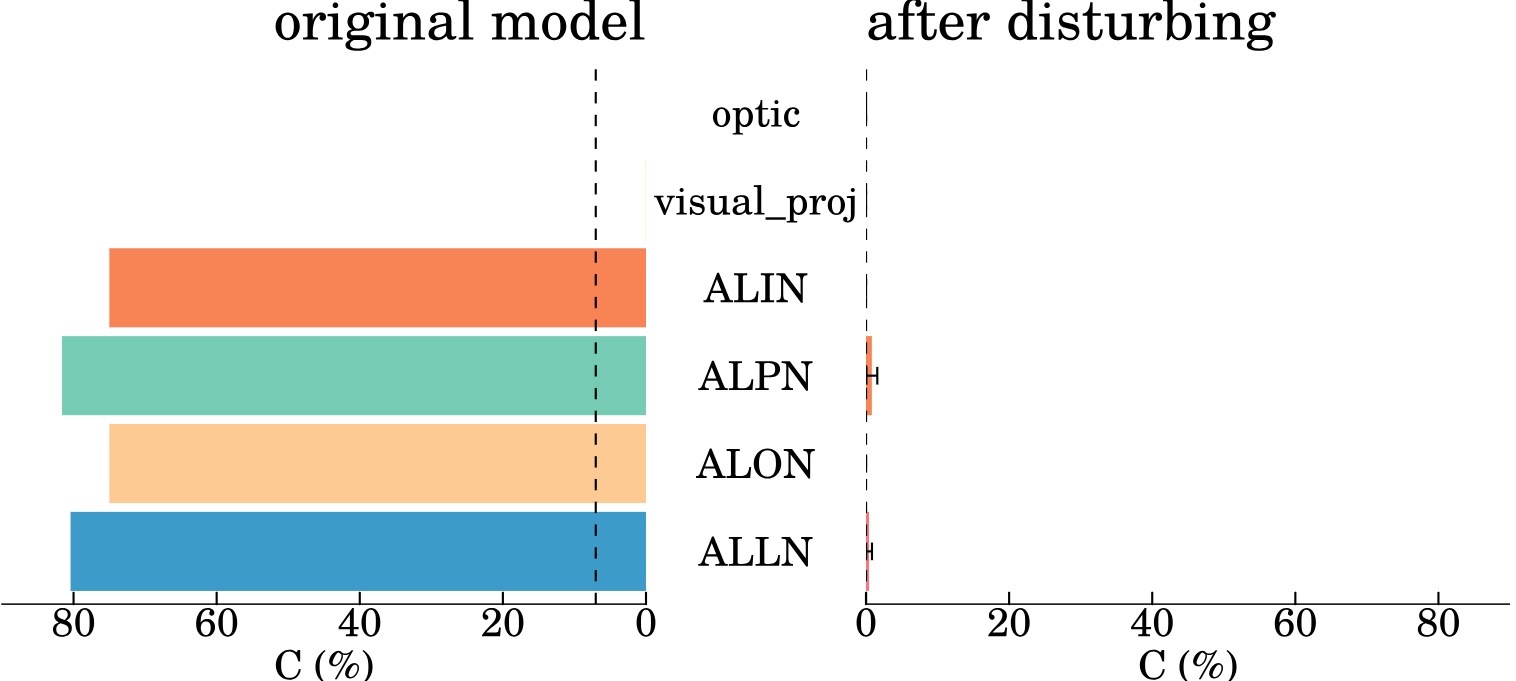}
			\end{minipage}		
		}
		\\
	\subfloat[LIF, visual]{
				\begin{minipage}[t]{0.5\linewidth}
				\includegraphics[width=7cm,height=3.5cm]{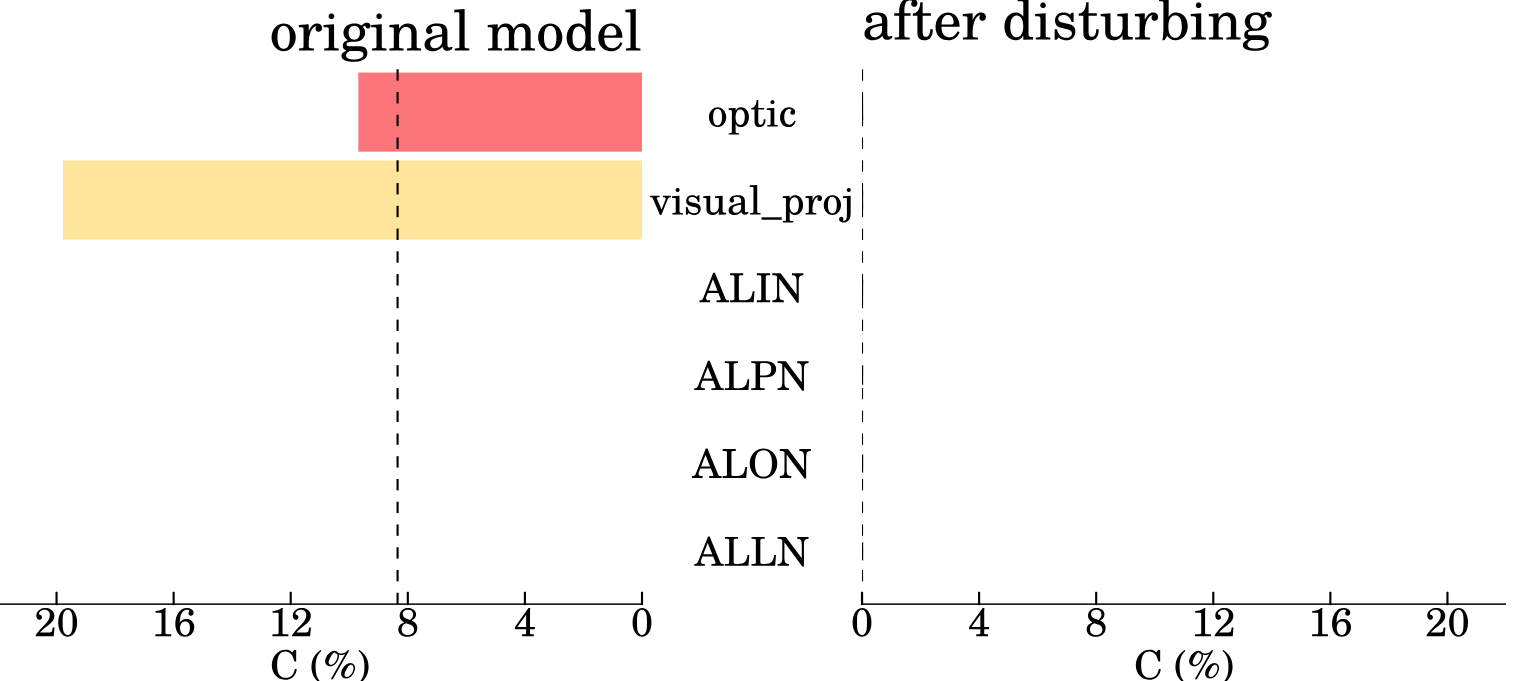}
			\end{minipage}		
		}
	\subfloat[LIF, olfactory]{
		\begin{minipage}[t]{0.5\linewidth}
				\includegraphics[width=7cm,height=3.5cm]{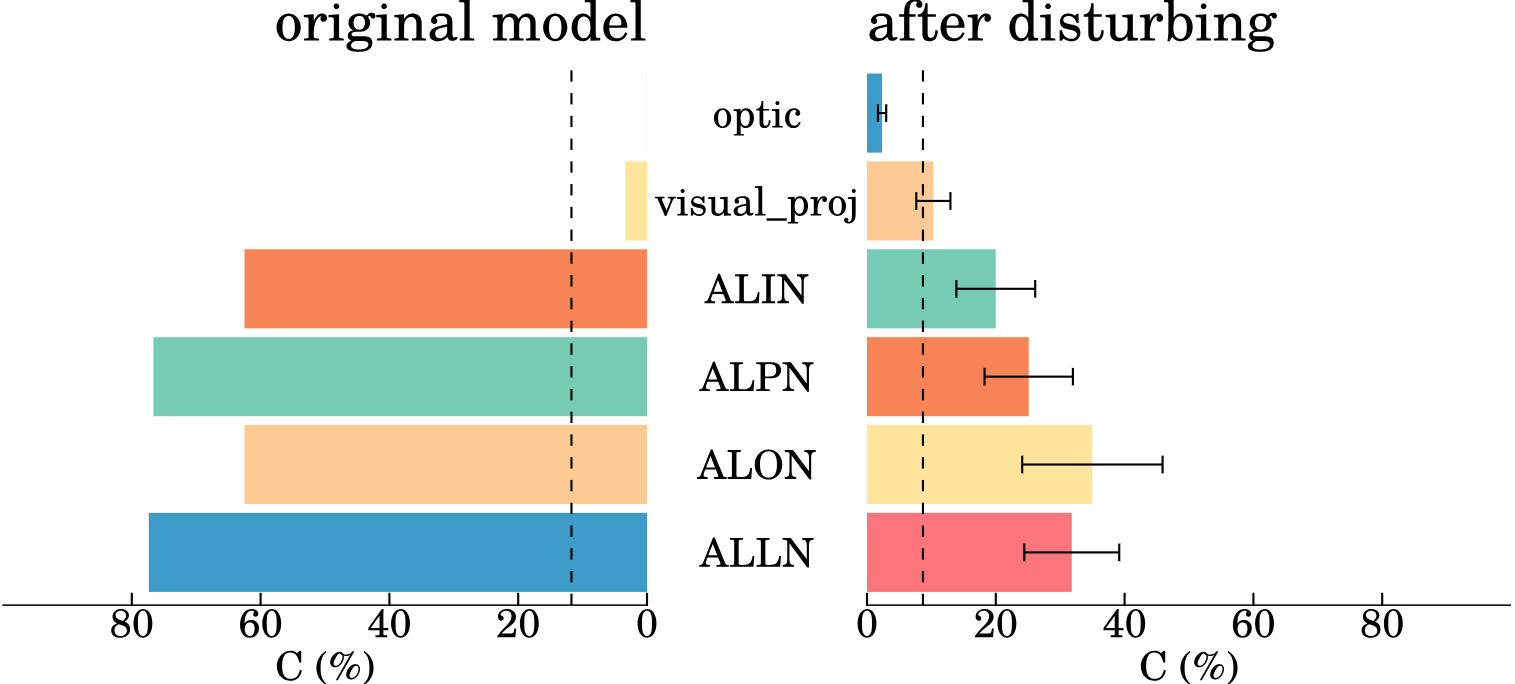}
			\end{minipage}		
		}
	\caption{Comparative test of intrinsic neurons' activation. For disturbing model, we conduct 5 repeated experiments and take the average and standard deviation. The dotted lines represent average activate rate in brain. In disturbing model, we set reconnect rate $p = 20\%$ and $\sigma$ = 0.8.}
	\label{scatter} 
\end{figure*}
\subsection{Separation of Visual and Olfactory Related Areas}
\begin{figure*}[t]
	\centering
	\subfloat[]{
	\begin{minipage}{0.5\linewidth}
	\includegraphics[width=8cm,height=7cm]{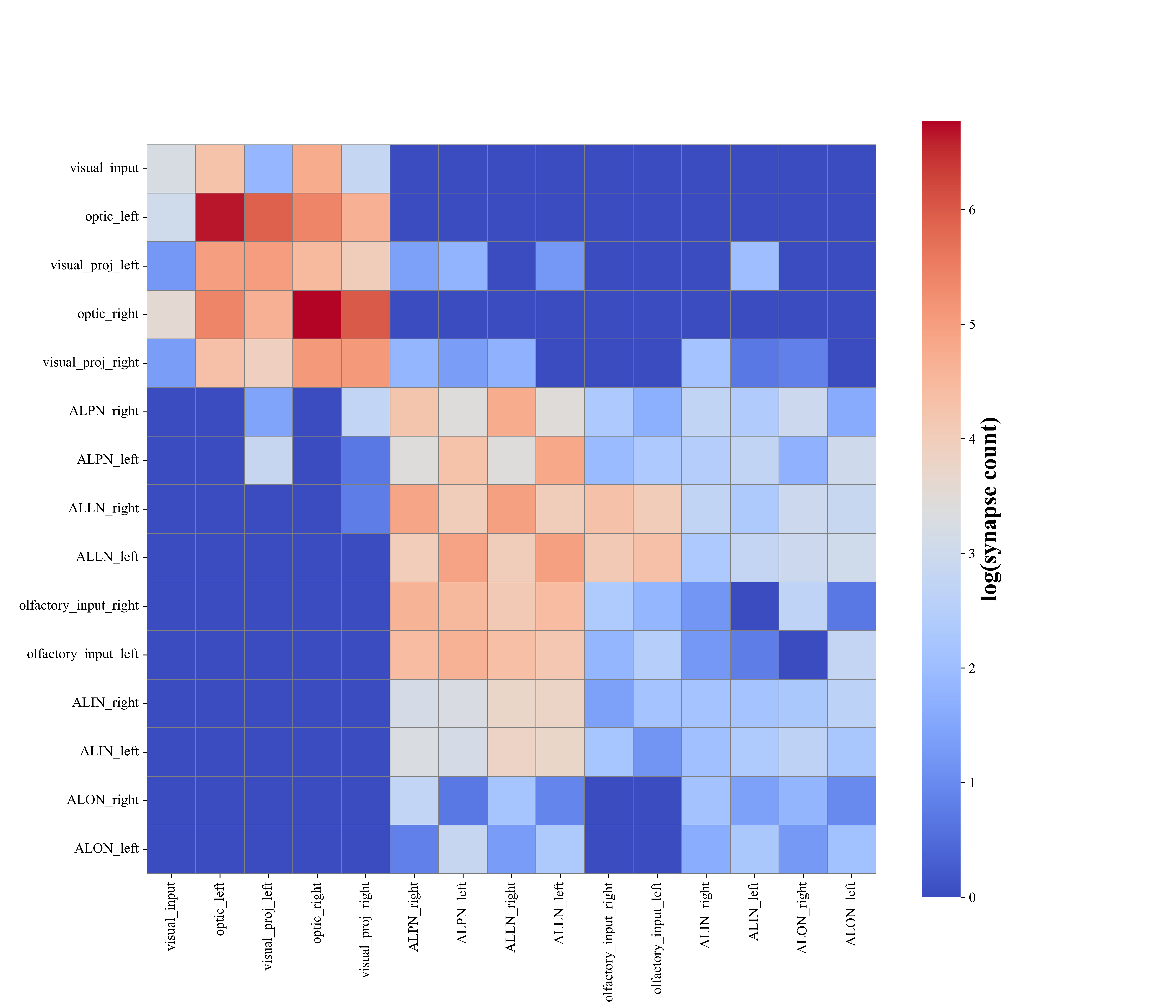}
	\end{minipage}
}
\subfloat[]{
	\begin{minipage}{0.5\linewidth}
	\includegraphics[width=9cm,height=7cm]{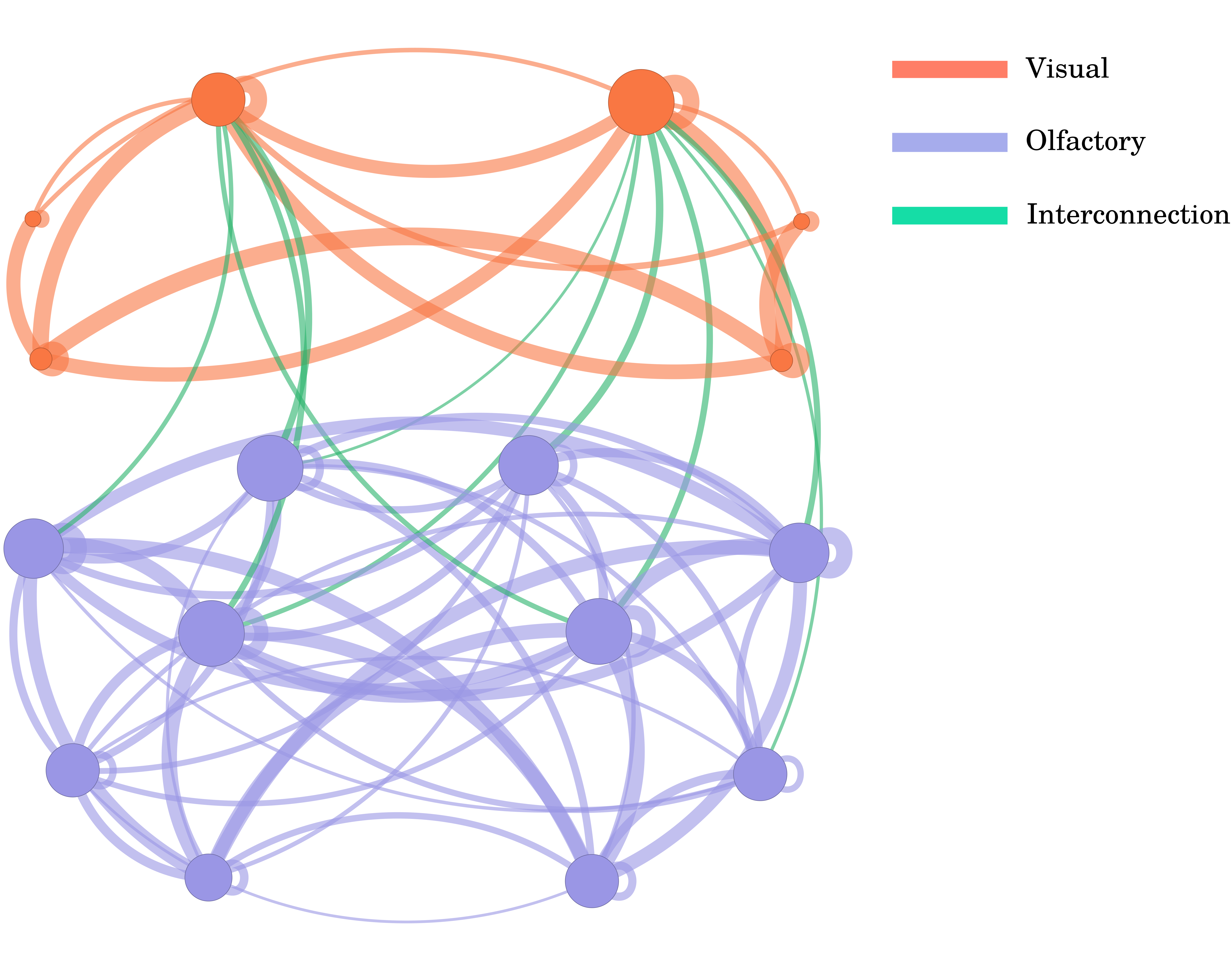}
	\end{minipage}
}
\caption{Heatmap and schematic diagram illustrating the connections between areas related to vision and olfaction. (a) presents the heatmap, while (b) shows the schematic diagram of the connection relationships.}
	\label{heat} 
\end{figure*}

To further investigate the causes of the above observed dynamic pattern that is independent of neuron models, we calculated the synapse count (number of connections) between various regions of the visual and olfactory systems and visualized the results as a heat map (Fig. \ref{heat}(a)). The findings reveal a significant segregation between the visual and olfactory areas. Specifically, within the visual and olfactory areas, there are 13,309,161 and 882,170 synapses, respectively. However, only 1,837 synapses exist between these two areas. These interconnecting synapses constitute only approximately 0.0138\% and 0.208\% of the total synapses within the visual and olfactory areas, respectively. 

Fig. \ref{heat}(b) represents the schematic diagram of the connection relationships in the neuronal visual and olfactory systems. We use nodes to represent functional areas and use edges to represent synapses between areas. The thickness of edges represents the number of connections. This diagram indicates that the number of connections within regions is significantly greater than the number of inter-regional connections. 

These indicate that the network structure between these two functional systems is almost entirely separate.  
\renewcommand\floatpagefraction{.99}
\renewcommand\topfraction{.99}
\renewcommand\bottomfraction{.99}
\renewcommand\textfraction{.1}
\setcounter{totalnumber}{50}
\setcounter{topnumber}{50}
\setcounter{bottomnumber}{50}
\subsection{Distances between Input and Activation Areas}
To further investigate whether this separation phenomenon is caused by greater spatial distance, we calculated the average network distance and spatial distance between different areas of the brain network. The results are presented in Tab. \ref{tab1}.

The analysis of Tab. \ref{tab1} showcases that spatial distances among neurons do not straightforwardly translate to ease of activation. While visual input neurons across all intrinsic areas exhibit similar average spatial distances, quasi-real activation patterns manifest predominantly in specific neural types (optic and visual\_projection). Conversely, the average network distances between input neurons and target neurons is notably smaller compared to irrelevant neurons. Specifically, visual input neurons exhibit network distances to optic and visual\_projection neurons as 5.7018 and 5.7432, respectively, while distances to neurons irrelevant to olfactory (ALLN, ALIN, ALON, and ALPN) are measured at 6.6442, 6.2350, 6.1506, and 6.5380, respectively. This demonstrates that neurons corresponding to visual inputs maintain a generally closer network distance relative to unrelated neurons. Furthermore, neurons associated with olfactory inputs maintain network distances approximately half that of unrelated neurons. These observations suggest that the network distance between input neurons and their specific activation areas is more significant than spatial distance, underscoring the importance of network structure over spatial structure in defining the quasi-real activation patterns of brain networks.

\begin{table*}[!th]  
	
	\centering  
	
		\vspace{-0.35cm} 
	\caption{Spatial and network distance between input areas and functional areas. }\label{tab1}  
	\begin{tabular}{c| c |c c c c c c }  
		
		\hline   
	Input type & Distance type & optic & visual\_projection & ALLN & ALON & ALPN & ALIN \\
		\hline   
		
		{\multirow{2}*{Visual input}} &Spatial distance (nm)& $365012 $ & $354725 $ & $351919 $ & $346925 $ &$348517$ &$348901$
		\\
		\cline{2-8}  
		&Network distance (edge)& $\mathbf{5.7018} $ & $ \mathbf{5.7432} $ & $ 6.6442 $ &$ 6.1506 $ &$6.5380$  & $ 6.2350 $  \\
		\hline  
		
		{\multirow{2}*{Olfactory input}} &Spatial distance (nm)& $302690 $ & $251144 $ & $110827 $ & $123179 $ &$115089$ & $173345$ \\
		\cline{2-8}  
		&Network distance (edge)& $ 5.7967 $ & $ 5.1706 $ & $ \mathbf{3.4458} $& $ \mathbf{3.6312} $ &$\mathbf{3.6820}$ & $ \mathbf{3.3845} $ \\
		\hline  
	\end{tabular}  
	
\end{table*}

\section{Discussion}

Our study, which utilized a large-scale network communication model derived from the Drosophila connectome, highlights the crucial importance of neural network architecture in generating brain-like activation patterns.

We first analyzed the structure of the Drosophila brain network, including the basic statistical properties and the symmetry of the network. An interesting phenomenon observed is the correlation between neuronal degree and geometric size: Fig. \ref{fig1}(f) demonstrates that the number of synapses formed by a neuron (represented by degree) increases logarithmically, rather than linearly, with respect to its surface area, volume, or length.
These characteristics of the brain can be explained by the Weber-Fechner’s law: for any sensory pattern, the perceived intensity is a logarithmic function of the physical intensity. Specifically, the size, length, and area changes can only cause a logarithmic change in the number of synapses the neuron has. These results suggest that although synaptic plasticity lead to a continuous increase in the number of synapses between neurons and other cells, this logarithmic relationship makes it difficult for the number of synapses formed by a neuron to increase indefinitely. This may account for the overall power-law distribution observed in Fig. \ref{fig1}(f). 
 
To examine the impact of neuronal activation mechanisms on the generation of realistic network patterns, we simulated three activation mechanisms for specific inputs.
Fig. \ref{scatter} suggests that, the real network structure is more critical than activation mechanisms in generating quasi-real activation patterns. The brain's unique network structure plays a crucial role in this realistic activation mechanism, aligning with studies on complex systems where the organization and interaction structure among individuals, rather than individual capabilities, determine emergent phenomena\cite{holland2000emergence}. 

To elucidate the influence of network structure on function,
we analyzed the topology of the network (Fig. \ref{heat} and Tab. \ref{tab1}). These results collectively suggest that network architecture, rather than spatial structure or activation mechanisms, is the primary determinant of the Drosophila connectome's ability to generate quasi-real activation patterns, owing to its unique connectivity profile.

To test the impact of network perturbation ratio on its functionality, we calculated activation patterns with different reconnect rates (Appendix \ref{sensitivity}). These results indicate that the formation of realistic activation patterns is highly sensitive to changes in network structure, with even minor perturbations capable of disrupting this network configuration. This indirectly substantiates that the authentic network structure is crucial for the formation of genuine activation patterns. Additionally, it highlights that complex functions (such as vision) are more susceptible to network disturbances. Conversely, Fig. \ref{single} shows that the network retains its original performance under a region-based attack. One explanation for these findings is that random rewiring thoroughly alters the network topology, potentially disrupting original functional modules and pathways, which leads to disordered information transmission paths and significantly impacts activation patterns. Additionally, because rewiring is global and random, it induces unpredictable perturbations that undermine redundancy and fault tolerance mechanisms, rendering the network more sensitive to such global disturbances. In contrast, the reason the network's dynamic properties remain relatively stable after the deletion of nodes within a specific region is that this perturbation is localized and has a minor impact on the global topology. Moreover, random rewiring may affect critical nodes or key pathways within the network, leading to significant functional impairment of the overall network. However, regional node deletion, if involving non-critical nodes and edges, may not significantly impact the network’s functionality, allowing it to maintain normal activation patterns. 

To evaluate the impact of the observed network structural symmetry (Fig. \ref{fig1}(a), (c), (d), and (e)), we input visual signals separately to the left and right hemispheres and observed the effects. The findings in Appendix \ref{Bi} reflect real-world phenomena: Although the left and right hemispheres possess certain functional specializations, they also exhibit synergistic effects and typically collaborate to fulfill specific functions \cite{swanson2012brain}. For example, during complex cognitive tasks, both hemispheres coordinate their efforts to achieve task completion. The experimental results indicate that similar synergistic effects emerge even when only the neurons on one side of the brain are stimulated, indirectly validating the model's effectiveness. Moreover, the network maintains similar synergistic effects even in the face of large-scale localized attacks, indirectly demonstrating the network's robust resilience. Finally, by analyzing the connections between different regions, it is found that the cause of these synergistic effects is not the connections between visual input and contralateral visual processing areas, but rather the extensive connections within the visual processing regions themselves. This provides a new perspective for understanding the brain's synergistic effects.



These findings has significant implications for artificial neural network research, suggesting that future AI designs may need to more closely mimic real brain network structures. Future research directions may include optimizing propagation rules and automatically adjusting link weights to achieve complete simulation of brain behavior. These advancements will pave the way for achieving genuine artificial intelligence while deepening our understanding of the nature of biological intelligence. 

\section{Acknowledgement}
We thank M.Murthy et al. for their invaluable help in sharing the Drosophila connectome data. This work was supported by the National Natural Science Foundation of China (72025405, 72088101, 72301285, 72001211), the National Social Science Foundation of China (22ZDA102), and the Natural Science Foundation of Hunan Province (2023JJ40685). The authors declare that they have no conflict of interest.

\bibliography{ref} 
\bibliographystyle{unsrt}
\clearpage
\appendix 
\textbf{\huge{APPENDIX}}
\setcounter{figure}{0}
\renewcommand{\thefigure}{S\arabic{figure}}

\section{Data and Methods}\label{method}

\subsection{Drosophila Connectome Data Set}
We utilize the recently released largest Drosophila brain connectome\cite{dorkenwald2023neuronal}, which provides a comprehensive mapping of the entire neural network within a female Drosophila brain. This data set includes over 120,000 neurons and 30,000,000 synapses, featuring detailed synaptic connections and high-precision 3D coordinates (accurate to the nanometer level) for both cell bodies and synapses. Additionally, it offers extensive labeling for various neuron types. For more detailed information, refer to \cite{dorkenwald2023neuronal,schlegel2023consensus}.

In this section, neurons are represented as nodes and synapses as connections between them. Tab. \ref{tab} presents the basic properties of the data set. We combine multiple edges between two nodes to calculate network statistics. The average degree before and after combining multiple edges is 501.6 and 37.01, respectively, indicating the prevalence of multiple synapses between pairs of neurons.

\begin{table*}[!htb]
	\centering
		\vspace{-0.35cm} 
	\caption{ Statistical Properties of Drosophila Brain Connectcome Graph}\label{tab3}
	
	\begin{tabular}{ c c c }
		
		\hline  \text { Number of Nodes} & \text {Number of Edges} & \text {Average Degree} \\
		
		$ 131459 $ & $ 32970606 $ & $ 37.01 $ \\
		\hline 
		\text {Average Degree (before combination)} &\text {Clustering Coefficient} &  \text {Eigenvector Centrality}\\
		$ 501.6 $ & $ 0.1527 $ & $ 0.0008817 $\\
		\hline
		\text {Network Diameter} &\text {Average Shortest Path} &  \text {Assortativity Coefficient}\\
		$ 11 $ & $3.9061 $ & $-0.05868$\\
		\hline
	\end{tabular}
	\label{tab}
\end{table*}

\subsection{Model}\label{Model}

In this section, we introduce the large scale network communication model which consumes less computational resource and can efficiently utilize the network structure. To solve the problem that the link weights between neurons can't be measured, we propose two hypotheses to estimate the link weights. In addition, we replace the node model in the network communication model with an improved LIF neuron model for comparison. 
\subsubsection{Symbol Systems}

\begin{figure}[!htbp]
	\centering
	\includegraphics[width=6cm,height=6cm]{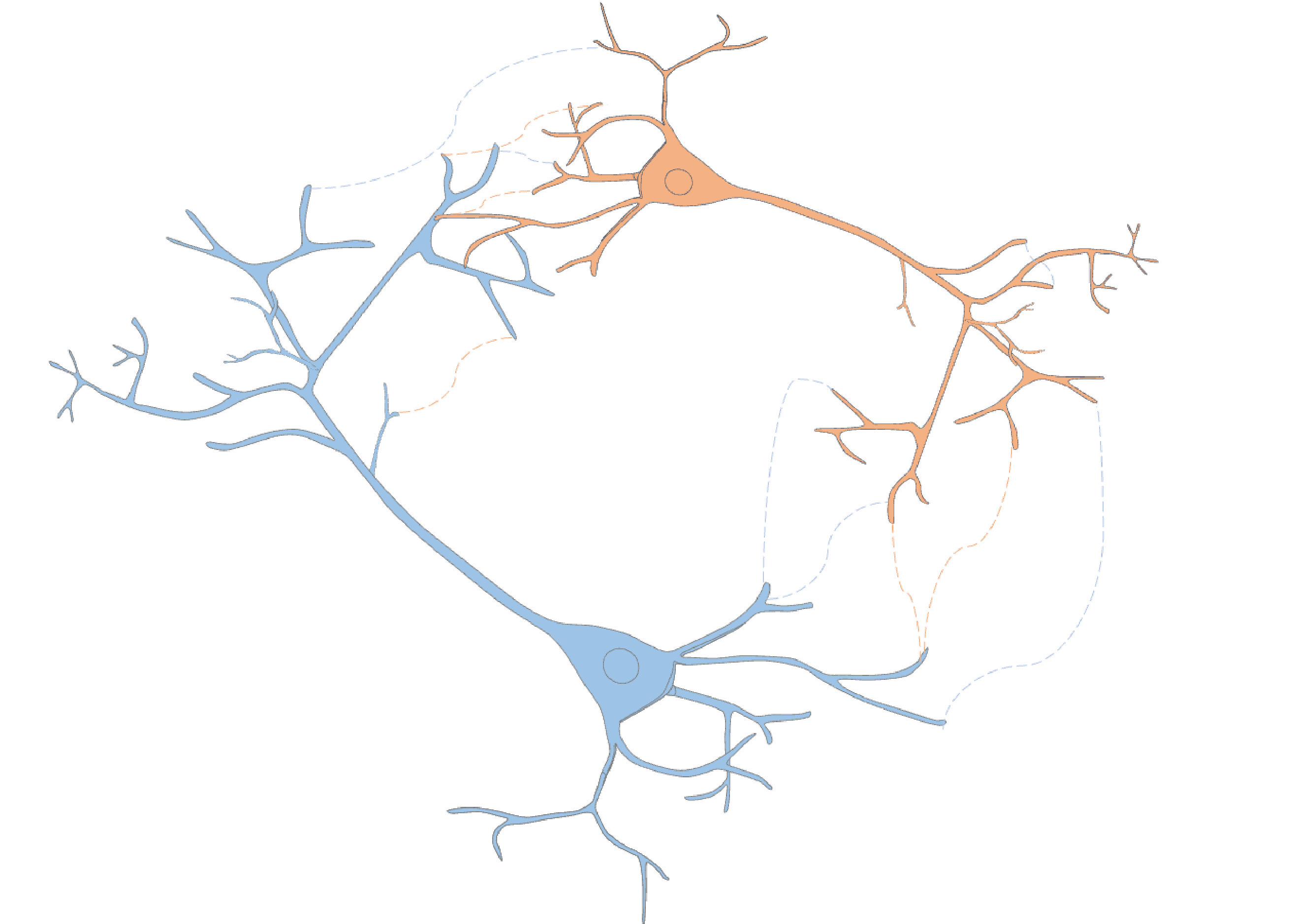}
	\caption{An example of complex connection}
	\label{complex}
\end{figure}

The original network representation of neural systems often uses nodes to represent neurons and edges to represent the links (chemical synapses, electrical synapses, etc.) between neurons\cite{avena2018communication}. However, this approach has several limitations: First, multiple synapses typically exist between two neurons\cite{fauth2015formation}, and these synapses are spatially and temporally asymmetric. For instance, two neurons, as illustrated in Fig. \ref{complex}, can have multiple synapses with varying dynamic properties. Additionally, due to differing spatial locations, signal transmission delays and the positions of action on downstream neurons also vary. Merging these synapses into a single edge oversimplifies the complexity of neural communication. 

For this reason, we use the dual method\cite{posfai2023impact} on Drosophila connectome data set to model the multi-link between neurons.  

\begin{figure*}[!]
	\centering
	\subfloat[original graph]{
		\begin{minipage}[t]{0.5\linewidth}
			\includegraphics[width=7cm]{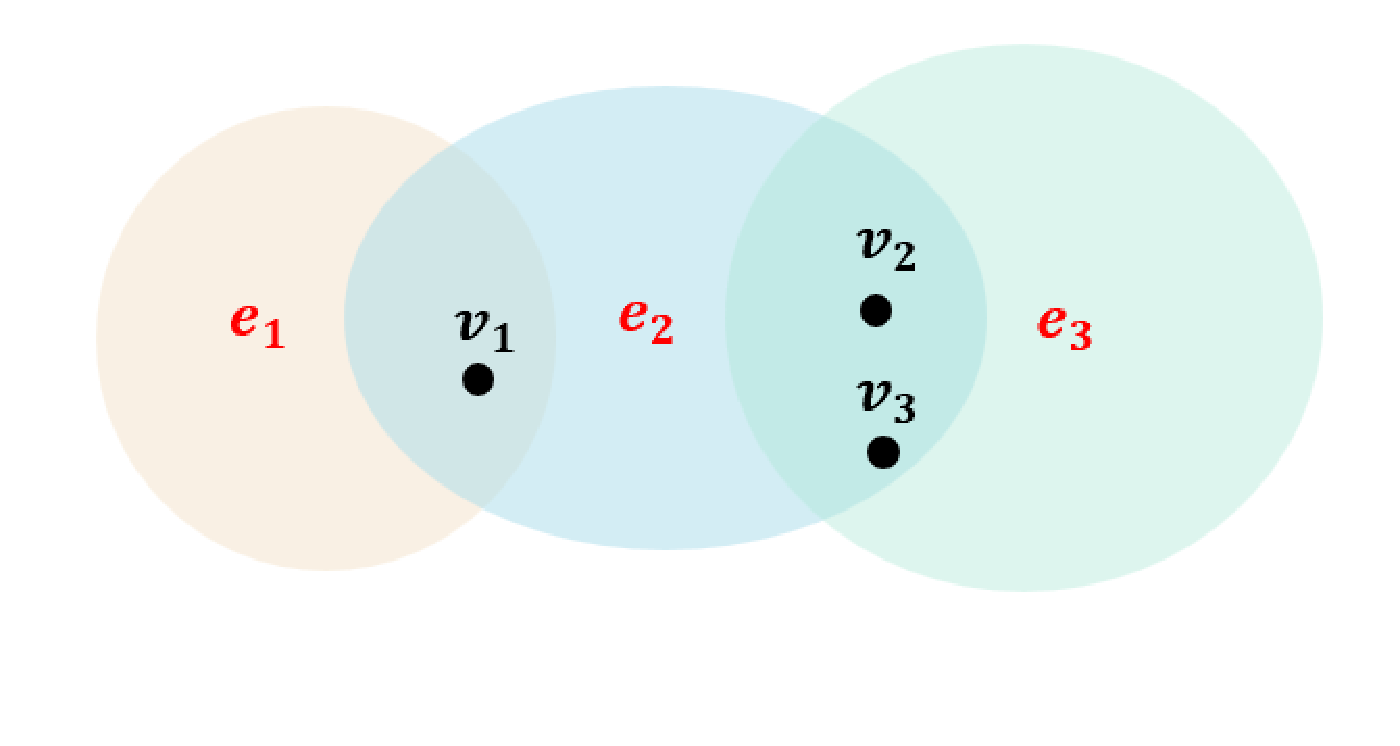}
		\end{minipage}		
	}
	\subfloat[dual graph]{
		
		\begin{minipage}[t]{0.5\linewidth}
			\includegraphics[width=9cm]{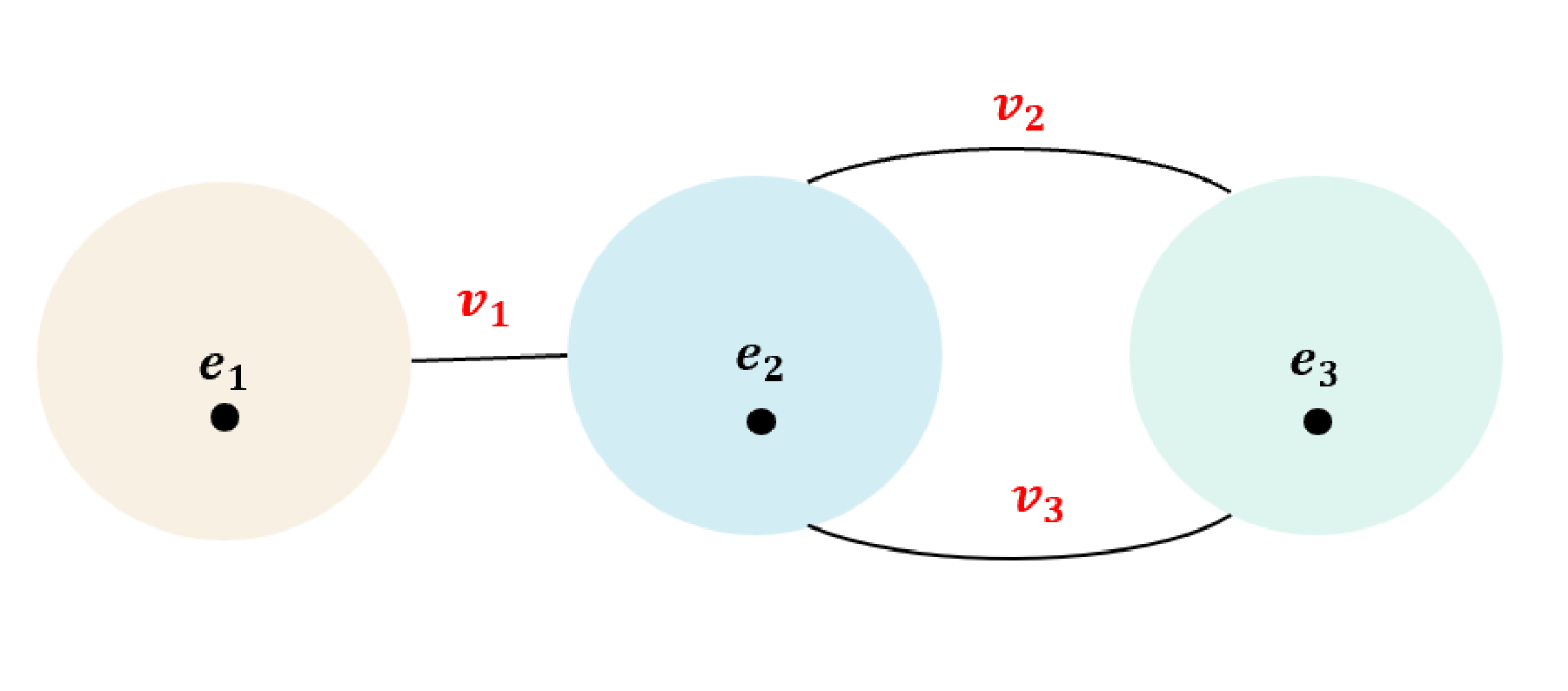}
		\end{minipage}	
		
	}
	
		%
		%
	\caption{An example of original graph and dual graph}
	\label{dual}
\end{figure*}

In a dual graph, the nodes and edges of the original graph are interchanged, resulting in a new graph. In simple terms, a dual graph is obtained by replacing the nodes of the original graph with edges, and the edges with nodes. Fig. \ref{dual} is an example of an original graph and a dual graph. After being processed, the nodes in the dualed graph represent the synapses and links represent the cell body. Then we can use the communication model to describe the message propagation process.

In order to standardize the symbol system, we use a edge (multi-edge, could include more than 2 nodes) to represent the neuronal cell body and use an node to represent a physical link between neurons (synapse). We regard the process of messages interactions as the nodes' interactions via edges. 

Let $N = \{n_1,...,n_i,...\}$ represents the set of all nodes and $E = \{e_1,...,e_j...\}$ represents the sets of all edges. An obvious relationship is formalized as the following manner. 
\begin{equation}
	e_k\cap e_q = \{n_i|(n_i \in e_k)\wedge (n_i \in e_q)\}.
\end{equation}
Note that the union of each element in $N$ and $E$ is the whole graph. The graph $G$'s definition is given as follows.
\begin{equation}
	G = \bigcup_{j}^{}e_j = \bigcup_{i}^{}\{n_i\}. 
\end{equation}
Let the label of any neuron be $j$. Because neural link is directed, let ${e_j}^+$ ( ${e_j}^-$) represent the set of elements which output (input) the signals, note that ${e_j}$ = ${e_j}^+ \cup  {e_j}^-$.

Each neuron and synapse has an activate state (state = 1) or resting state (state = 0). When a neuron is at the activate state, it fires to other neurons, otherwise it keeps rest. So set the statement of the element $i$ as $X_i$. $X_i = 1$ if the element $n^{}_i$ is active, otherwise $X_i = 0$. Then let $Y_{j}$ represent the statement of set $e_j$. $Y_{j} = 1$ if the set $e_j$ is active, otherwise $Y_{j} = 0$. 
\subsubsection{Large Scale Network Communication Model (LSNC)}
By modeling the Drosophila connectome as graph, we introduce the large scale network communication model on this graph. The general transmission function is defined in the following manner,

\begin{equation}
	\label{15}
	Y_j= \sigma(\frac{1}{\beta_j}\sum_{i|e_{i}^{+}\cap e_{j}^{-}\ne \emptyset}^{} \sum_{k|n_k\in e_{i}^{+}\cap e_{j}^{-}}\omega (n_k,e_i,e_j)X_k) ,
\end{equation}
where $\omega (n_k,e_i,e_j)$ represents the link weight between set $i$ and set $j$ (specifically, two neurons) through element $n_k$. Here we consider that the link weight between two neurons is not only related to the neurons themselves, but also to the synapses between them. $\sigma(.)$ is the activate function. $\beta_j$ is the normalization factor. 

In this paper we use threshold activation function and sigmoid function for comparison.

The threshold activation function is a simple, binary decision function used in artificial neurons. It outputs a value of 1 if the input exceeds a certain threshold, and 0 otherwise. This function is also known as the Heaviside step function.
\begin{equation}
	\sigma(x)=\left\{
	\begin{aligned}
		1 \quad x\ge threshold\\
		0 \quad x < threshold\\
	\end{aligned}
	\right
	.
\end{equation}
The sigmoid activation function is a smooth, S-shaped function that maps input values to an output range between 0 and 1. It is commonly used in neural networks to introduce non-linearity and to model probability-like outputs in binary classification problems. The sigmoid function is defined as the logistic function.
\begin{equation}
	\sigma (x) = \frac{1}{1+e^{-x} } 
\end{equation}
The statement of an neuron will influence the statement of all output synapses. The equation between neuron $j$ and output synapse $i$ is as the following manner. 

\begin{equation}
	X_i = Y_j \ \ \ \ if \ n_i \in e^{+}_j .
\end{equation}

The link weights between neurons determine the extent of influence they have on each other. Larger link weights mean a tighter "link" between two neurons. In everyday life, the brain continuously changes the link weights between neurons in response to inputs from the external world, allowing intelligence to emerge from the micro to the macro. Therefore, if researchers want to simulate the process of information propagation in the brain on large scale in a computer, determining the link weights between neurons is essential. However, existing experimental methods are not well-suited for directly measuring the weights between neurons in a whole brain, so adopting some alternative approach is necessary. Below, we introduce a hypothesis that use different methods to estimate the connection weights between neurons.

As previously mentioned, there exists a large number of synapses between two neurons. However, in practical applications, it is difficult to set the parameters for these edges individually. Consequently, a simple assumption naturally arises: assuming that the source neuron has the same influence on the target neuron through every synapse. We propose an assumption that every synapse is equivalent, which is widely used in neuron graph statistics analysis\cite{lin2023network}. That is to say, 
\begin{equation}
	\omega (n_k,e_i,e_j) = 1.
\end{equation}

Additionally, We denote the sum of all link weights between neurons \( i \) and \( j \) as $W_{ij}$:
\begin{equation}
	W_{ij} = \sum_{k|n_k\in e_{i}^{+}\cap e_{j}^{-}}^{} \omega (n_k,e_i,e_j).
\end{equation}
And we define $\beta_{j}$ as the maximum weight among all $W_{ij}$ related to neuron $j$. That is to say,
\begin{equation}
	\beta_{j} = { {Max}(\{ W_{ij}| e_{i}^{+}\cap e_{j}^{-}\ne \emptyset}\}).
	\label{19}
\end{equation}
Where $Max(S)$ represents the Maximal element in set $S$.  

Then the normalized equation is as follows.
\begin{equation}
	Y_j= \sigma(\frac{1}{\beta_j}\sum_{i|e_{i}^{+}\cap e_{j}^{-}\ne \emptyset}^{} \sum_{k|n_k\in e_{i}^{+}\cap e_{j}^{-}}\omega (n_k,e_i,e_j)X_k ) .
\end{equation}

\subsubsection{Neuron Dynamic Model}
To better reflect real-world conditions and demonstrate that the superiority of this network architecture does not solely rely on the design of the activation function, we also employed a variant of the LIF model\cite{lansky2008review} as the basic neuron model.

The membrane potential change of a neuron's voltage can be described by the following differential equation.  
\begin{equation}
	V_j(t+dt) = V_j(t) + \left( \frac{dt}{\tau} \right) \left( R \cdot I^{syn}_j - (V_j(t) - V_{\text{reset}}) \right)
\end{equation}
Where $V_j$ represents the membrane potential of $j^{th}$ neuron, $R$ represents the resistance constant and $ I^{syn}_j$ represents the total input current of neuron $j$. $\tau$ represents the time constant.

The input current $I^{\text{syn}}_j$ can't be calculated directly. To estimate the value of $I^{\text{syn}}$, we make the following assumptions based on the following supposes:

1. The magnitude of the current on the synapse should be proportional to the voltage difference between the connected neurons. 

2. The current on each synapse should be proportional to the connection weight between the neurons.

3. A current can only be emitted if the output neuron is in an active state.

Considering the supposes above, the equation of input current $I^{\text{syn}}_j$ is designed as follows:

\begin{equation}
	I^{\text{syn}}_j = \frac{1}{\alpha\beta_j}\sum_{_{i|e_{i}^{+}\cap e_{j}^{-}\ne \emptyset}^{}}\sum_{k|n_k\in e_{i}^{+}\cap e_{j}^{-}} \left( \omega (n_k,e_i,e_j) \times (V_i^{}-V_j) \times Y_i \right)  , 
\end{equation}
where $V_i-V_j$ represents the voltage difference between neuron $i$ and $j$. $Y_i$ represents the state of output neuron $i$. $\beta_j$ is defined according to equation (\ref{19}). $\alpha$ is the proportion const. 

According to the LIF model, when a neuron's voltage reaches the threshold value, the neuron will fire and then quickly resets its voltage. Based on this, the mechanism for updating the neuron voltage is designed as follows.

\begin{equation}
	\begin{aligned}
		\text{if} \ V_j \geq V_{th}  & : \ Y_j = 1 \ \text{and} \ V_j = V_{\text{rest}}
		\\
		\text{else}  & : \  Y_j = 0
	\end{aligned},
\end{equation}
where $V_{th}$ represents the fire voltage and $V_{\text{rest}}$ represents the rest voltage, respectively. 

Other propagation mechanism is consistent with network model introduced in section B.2.2. 
\subsection{Evaluation Criteria}
In previous study, researchers used classic indicators such as signaling cost and activation time to evaluate the quality of the model\cite{seguin2023brain}\cite{kaiser2007criticality}\cite{seguin2023communication}. However, these indicators can't describe the model accords with the true communication process or not, even if the activation time is short and the signaling cost is minimum, etc. To better assess whether a model can reflect the real situation of brain communication, we use the percentage of activated neurons in regions that related to (should work under) a certain stimulus. To illustrate, if the brain gets a visual stimulus, the region related to vision should work while the region not related to vision should remain a resting state. Then we record the percentage of neurons activated. If regions related to the certain function have a significant higher activation percentage than the average activation percentage of the whole brain, we call it a quasi-real activation pattern. 

We define the consistency indicator $C$ in a area which has a certain function (vision, olfaction, etc) is defined in the following manner.
\begin{equation}
	C =  N^{act}/N , 
	\label{Criteria}
\end{equation}
where $N^{act}$ represents the number of neurons activated in this area at the final state,  $N$ represent the number of neurons in this area and $N$ represent the total number of neurons.

\section{Introduction of Related Neurons}\label{neuron}
\emph{Optic neurons}, also known as retinal ganglion cells, are the first neurons in the visual pathway. They receive signals from the photoreceptor cells in the retina and transmit this information via the optic nerve to the brain. These neurons play a crucial role in carrying visual stimuli from the eye to the brain, where further processing and interpretation of the visual information occur. 

\emph{Visual projection neurons} are a type of neuron found in the visual system that transmit visual information from one region of the brain to another. They play a crucial role in processing and relaying visual signals from the retina to various visual centers in the brain, including the thalamus, hypothalamus, and primary visual cortex (V1), among others.

\emph{Antennal Lobe Input Neurons} (ALINs) are primarily olfactory sensory neurons (OSNs) that carry olfactory information from the sensory structures, such as the antennae and maxillary palps, directly to the antennal lobe. Each OSN expresses a specific odorant receptor and responds to particular chemical cues. The axons of these neurons synapse with projection neurons and local neurons within the glomeruli of the antennal lobe.

\emph{Antennal Lobe Projection Neurons} (ALPNs) transmit the olfactory information in the antennal lobe. These neurons receive input from the olfactory sensory neurons within the glomeruli and project their axons to other brain areas, such as the mushroom body and the lateral horn, where further processing and integration of olfactory information occur.

\emph{Antennal Lobe Output Neurons} (ALONs) refer to the projection neurons that serve as the primary output pathways of the antennal lobe, sending processed olfactory information to higher brain regions.

\emph{Antennal Lobe Local Neurons} (ALLNs) are inter neurons that are confined to the antennal lobe. They typically have processes (dendrites and axons) that branch extensively within the antennal lobe and form connections with multiple glomeruli. These neurons play a critical role in modulating and refining the olfactory information by providing inhibitory or excitatory input to projection neurons and other local neurons, thus shaping the olfactory responses and contributing to odor discrimination and perception.

Tab. \ref{num} represents the numbers of specific neurons of each hemisphere in related areas. It presents that the number of neurons in specific areas of the Drosophila brain connectome is mirror-symmetrical. Additionally, visual neurons constitute the largest proportion among the connectome's neurons (approximately one-twice). Neurons in the olfactory-related regions, though comprising a smaller proportion (for instance, the ALON region has only 8 neurons on each side), play a critical role in olfactory perception.

\begin{table*}[!htb]
	\centering
		\vspace{-0.35cm} 
	\caption{Node numbers in related areas Drosophila Brain Connectcome Graph}
	
	\begin{tabular}{ c c c c}
		
		\hline  \text {total (left)} & \text {total (right)} & \text {optic (left)}& \text {optic (right)} \\
		
		$ 60834$ & $ 63144$ & $36131 $& $ 37524 $ \\
		\hline 
		\text {visual projection (left)} & \text {visual projection (right)}& \text {ALIN (left)} & \text {ALIN (right)}\\
		$3117 $& $ 3117 $ & $ 12 $& $12 $\\
		\hline
		\text {ALLN (left)} & \text {ALLN (right)}& \text {ALON (left)} & \text {ALON (right)}\\
		$ 196 $ & $194 $ & $8$ & $8$\\
		\hline
		\text {} & \text {ALPN (left)}& \text {ALPN (right)} & \text {}\\
		$  $ & $310 $ & $314$ & $ $\\
		\hline
	\end{tabular}
	\label{num}
\end{table*}
%
%
%

For further information, please read \cite{schlegel2023consensus}.
\section{Experimental Settings}

The basic methodology of the experiment is as follows. We apply a continuous stimulus signal (binary 0-1 signal) to a subset of input neurons (i.e., visual or olfactory neurons) in the brain. We then compute the signal's propagation process within the brain and record the responses of both intermediate and output neurons. We give 2 types of input neurons (visual, olfactory) stimulus and record the activation patterns of 6 type of neurons: optic (visual), visual\_projection (visual), ALLN (olfactory), ALIN (olfactory), ALPN (olfactory) and ALON (olfactory). When we give neuron a stimulus, we set the state of neuron as 1. The evaluation criterion involves calculating the ratio of neurons that should be activated to those that are actually activated in a specific area (see Equ. \ref{Criteria} in Appendix \ref{method}).

Given that the brain functions similarly to a shallow neural network\cite{suzuki2023deep}, we document outcomes at an iteration step size of 5. Since the input is constant and no randomization parameters are introduced, the results of each experiment are consistent, eliminating the need for repetition. For detailed information on related neurons, please read Appendix \ref{neuron}.



In the linear threshold model, we set $\sigma_0 = 0.8$. For the simulation of the LIF model, numerous hyperparameters are required. All the hyperparameters used are listed in the following table. Biological hyperparameters are set according to real biological experimental tests\cite{peron2009spike}. The current proportion constant is the optimal parameter obtained through equidistant search.

\begin{table}[h]
	\centering
	\caption{List of Hyperparameters}
	\begin{tabular}{lll}
		\toprule
		Hyperparameter & Description & Value \\
		\midrule
		$\tau$ & Membrane time constant & 15.0 ms \\
		$V_\text{rest}$ & Resting membrane potential & -65 mV \\
		$V_\text{th}$ & Spike threshold & -45 mV \\
		$R$ & Membrane resistance & 0.5G$\Omega$ \\
		$dt$ &time step& 0.55 ms\\
		$\alpha$ & Current proportion const & 0.03/0.045 (visual/olfactory) \\
		\bottomrule
	\end{tabular}
\end{table}
To demonstrate the importance of network structure, we add a comparative model that randomly reconnects networks. Specifically, we perform degree-preserving rewiring on a certain proportion of edges in the brain, which means perturbing the network structure while ensuring that the degree of each node remains unchanged. In Fig. \ref{scatter} we set reconnect rate $p = 20\%$. For its sensitivity analysis, please read Appendix \ref{sen}.

\section{Bilateral Response under Unilateral Stimulus}\label{Bi}
In real life, the left and right hemispheres cooperate to perform complex tasks. To examine whether network models can simulate this cooperation, we administered a unilateral stimulus to input neurons and monitored the varying activation ratios across both hemispheres. Specifically, we independently simulated visual input neurons in the left and right hemispheres and recorded the rate of activated neurons in each hemisphere over time.
Second, we investigate whether local attack to one side of the network—analogous to real-world conditions such as tumors and other diseases—affects this bilateral response process, and how the extent of such damage influences the synergistic response. 

The specific methods of inducing attack are as follows: We first identify two symmetrical coordinates on both hemispheres of the brain. Using these coordinates as the center, we remove nodes within varying radii and subsequently record the response curves for each hemisphere-related region. 

\begin{figure*}[!thbp]
	\centering
	\subfloat[Geometric dimensions of Drosophila connectome]{
		\begin{minipage}[t]{0.48\linewidth}
			\includegraphics[width=8cm,height=4cm]{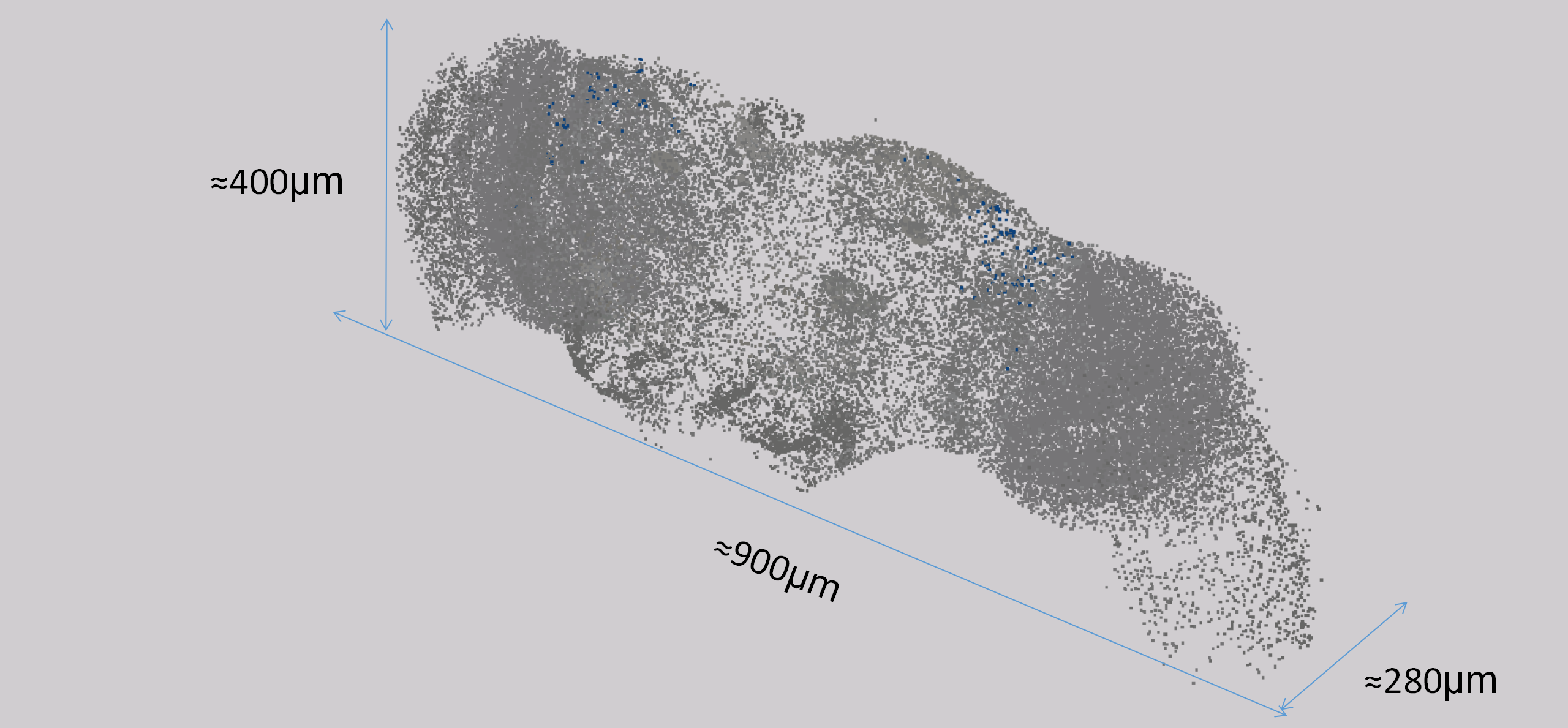}
		\end{minipage}		
		
	}
	\subfloat[Locations of targeted regions]{
		\begin{minipage}[t]{0.48\linewidth}
			\includegraphics[width=8cm,height=4cm]{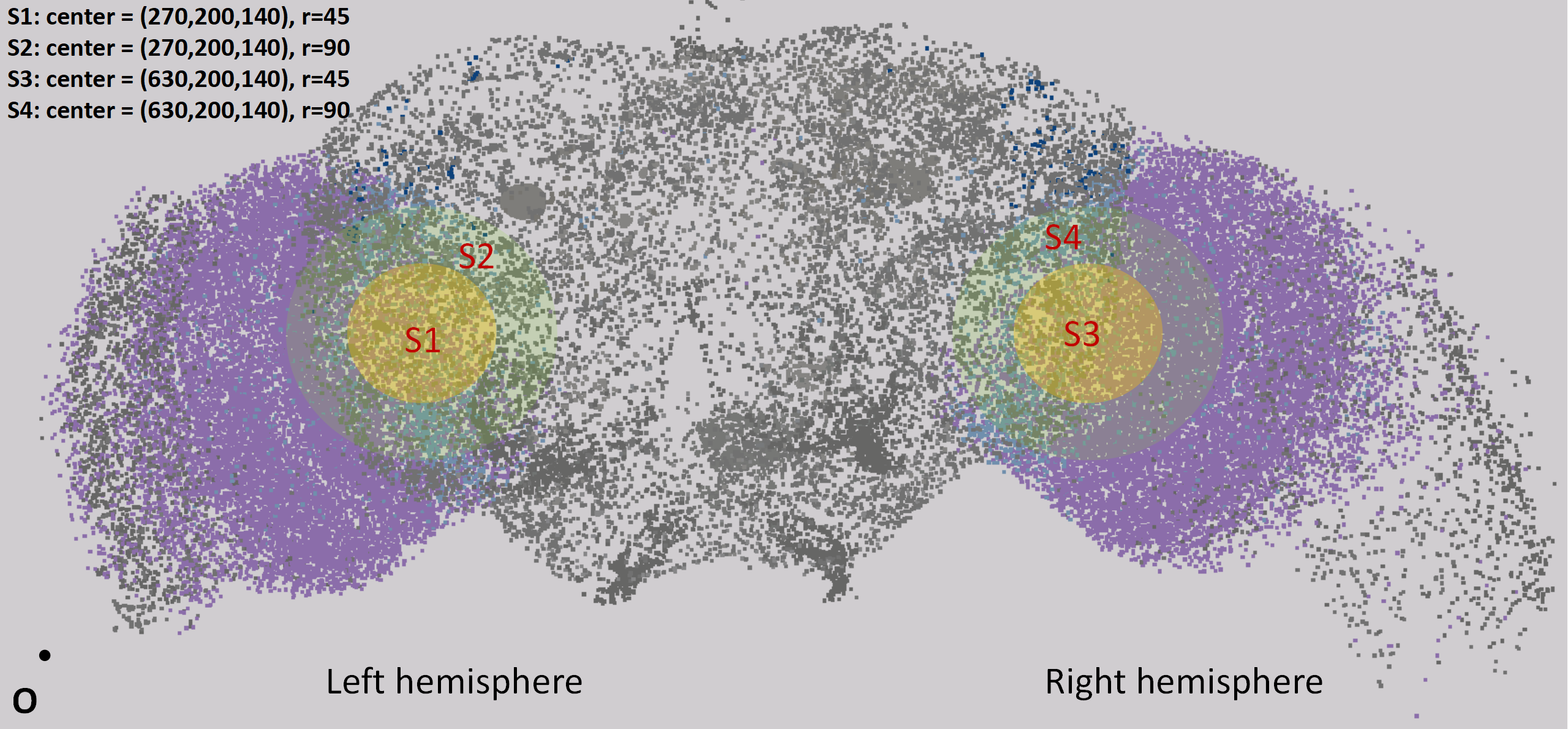}
		\end{minipage}		
	}
	\caption{Schematic diagram of the specific length measurements of the Drosophila connectome and the location of regions deleted. The units of center and r in labels are micrometers. }
	 \label{schematic}
\end{figure*}

Fig. \ref{schematic} illustrates the specific length measurements of the Drosophila connectome and the regions examined in this study. The purple and blue cells represent neurons associated with visual processing. Spheres $S1$, $S2$, $S3$, and $S4$ each represent four spherical regions subjected to localized attacks.
 Sphere $S1$ and $S2$ encompass approximately 2,000 neurons each, while sphere $S3$ and $S4$ each cover approximately 12,000 neurons.

\begin{figure}[!thbp]
	\centering
	\subfloat[Left stimulus]{
		\begin{minipage}[t]{0.5\linewidth}
			\includegraphics[width=9cm,height=7.2cm]{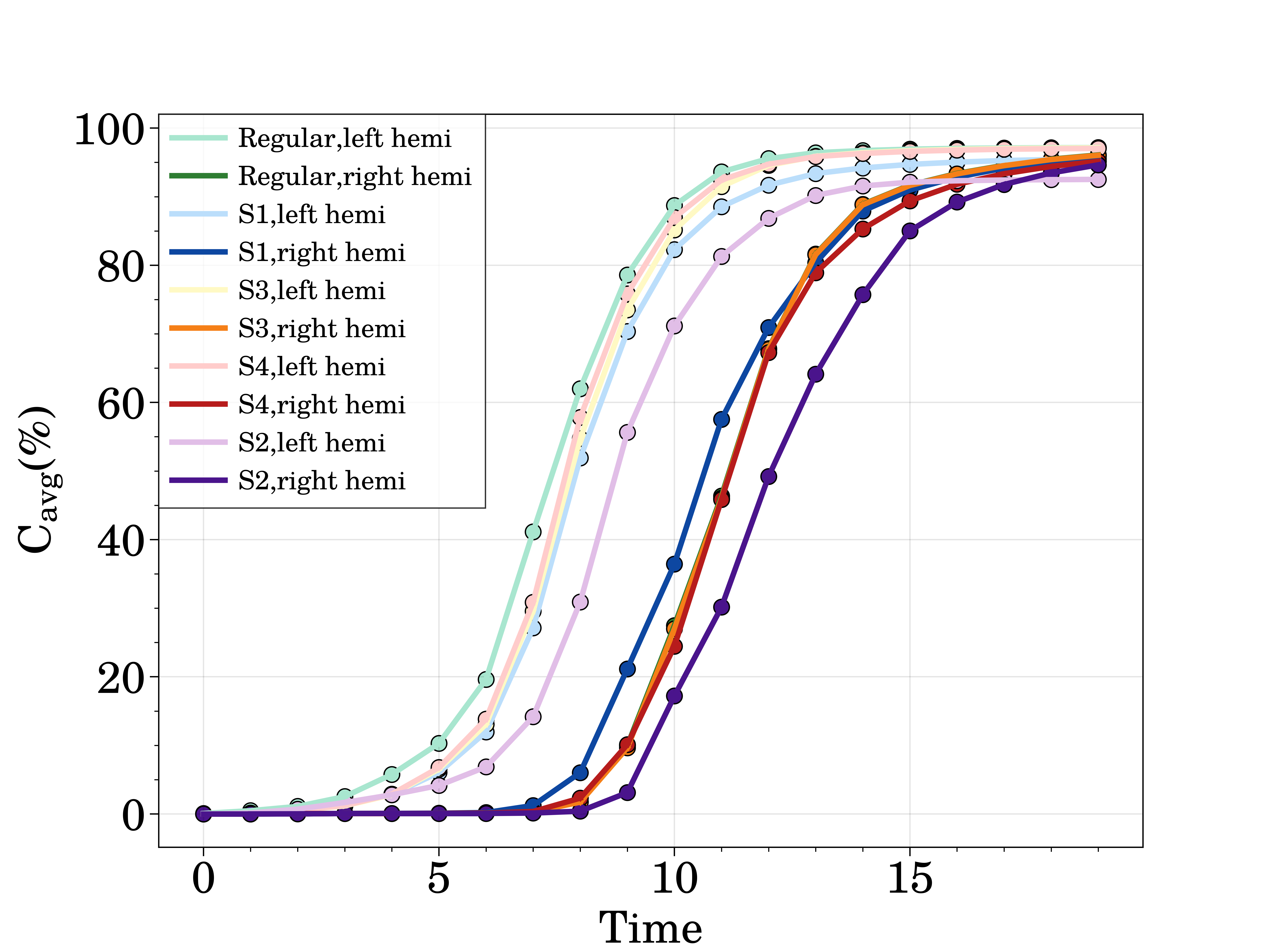}
		\end{minipage}		
		
	}
	\subfloat[Right stimulus]{
		\begin{minipage}[t]{0.5\linewidth}
			\includegraphics[width=9cm,height=7.2cm]{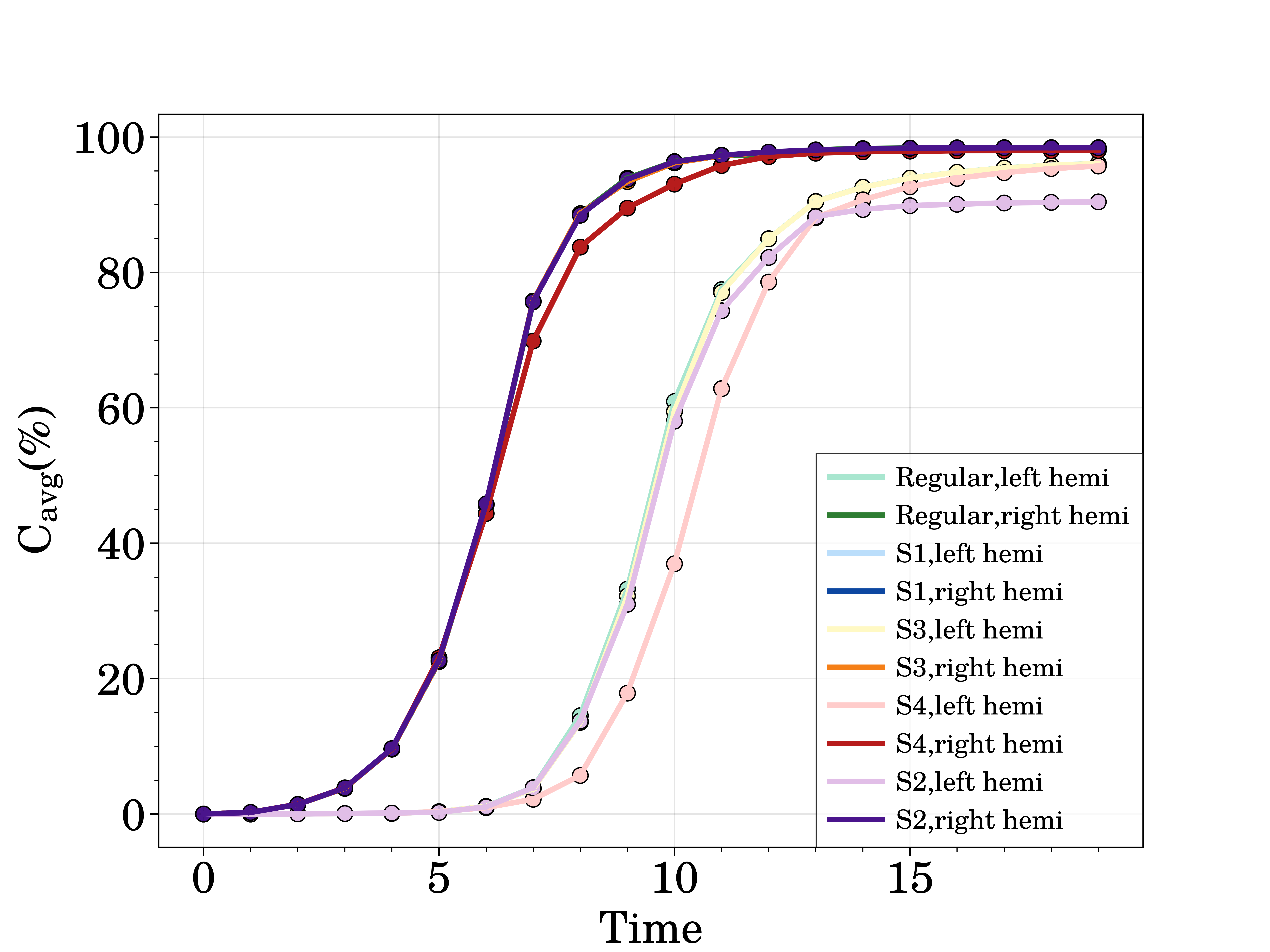}
		\end{minipage}		
	}
	\caption{Experiment of activation rate under single stimulus. All responses on the left hemisphere are marked with light-colored lines, and all responses on the right hemisphere are marked with dark-colored lines. We set threshold $\sigma_0 = 0.95$. 
	}
	\label{single}
\end{figure}

Fig. \ref{single} illustrates the curve depicting how the response ratio evolves with each time step. We use $C_{avg}$ to represent the average activated rate of areas related to vision. Notably, when stimulating one hemisphere alone, visual processing neurons in both hemispheres are activated (see curve regular left and right in Fig. \ref{single}(a) and (b)). 

Secondly, in the case of different regional node failures, even when roughly 12,000 nodes (approximately 1/10 of the total network nodes and 1/3 of the visually relevant neurons) were deleted, the network ultimately exhibited response characteristics similar to the original curve. This indicates that the network possesses strong robustness facing with local attacks. Additionally, for networks with failures on the input side, a delay in bilateral response curves was observed when the failure radius reached 90$\mu m$  ($S4$, left and right stimuli). This suggests that large-scale failures of visually relevant nodes primarily affect the rate at which the visual functional network reaches a steady state, rather than the number of nodes activated at steady state.

Below, we further explain the reasons for the significant delay in responses to unilateral large-scale regional attacks on this side's stimuli (e.g., in Fig. \ref{single}(a), where the local damaged area is the $S2$ region of the left hemisphere, and stimuli on the left side cause responses in both the left and right hemispheres (light purple and dark purple curves) to be delayed). Calculations show that there are 19,031 synapses from the left hemisphere's visual input area to the left hemisphere's visual signal processing area (optic, visual\_projection), while there are only 18 synapses from the left hemisphere's visual input area to the right hemisphere's visual signal processing area. Furthermore, there are 338,456 synapses from the left hemisphere's visual processing area to the right hemisphere's visual signal processing area. This implies that, primarily, the unilateral visual input area essentially does not interact with the contralateral visual signal processing area, and interactions occur indirectly through the ipsilateral visual processing area. Consequently, when the visual signal processing area of this side is damaged, the reduction in the number of neurons makes it challenging for the remaining neurons to reach an excitatory state, causing delays in the ipsilateral visual signal processing area. This delay, in turn, leads to delayed responses in the contralateral visual signal processing area.

%
%
\section{Impact of Perturbation Rate on Activation Areas}\label{sen}
In this section, we discuss the impact of different reconnection rates on activation patterns using the degree-preserving reconnection method.

The experiments presented in this section involve calculating patterns induced by two types of stimuli across three models under network perturbations of 1\textperthousand, 5\textperthousand, 1\%, 3\%, 5\%, 10\%, and 20\% reconnect rates. Let $C$ be the average activation ratio of an area. We normalized each data point using the difference between the activation ratio of this area and the average activation ratio $\bar{C} $ of the entire brain, that is to say, $\Delta C = C - \bar{C} $.

\begin{figure}[!thbp]
	\centering
	\subfloat[threshold model, visual stimulus]{
		\begin{minipage}[t]{0.33\linewidth}
			\includegraphics[width=5.5cm,height=4.8cm]{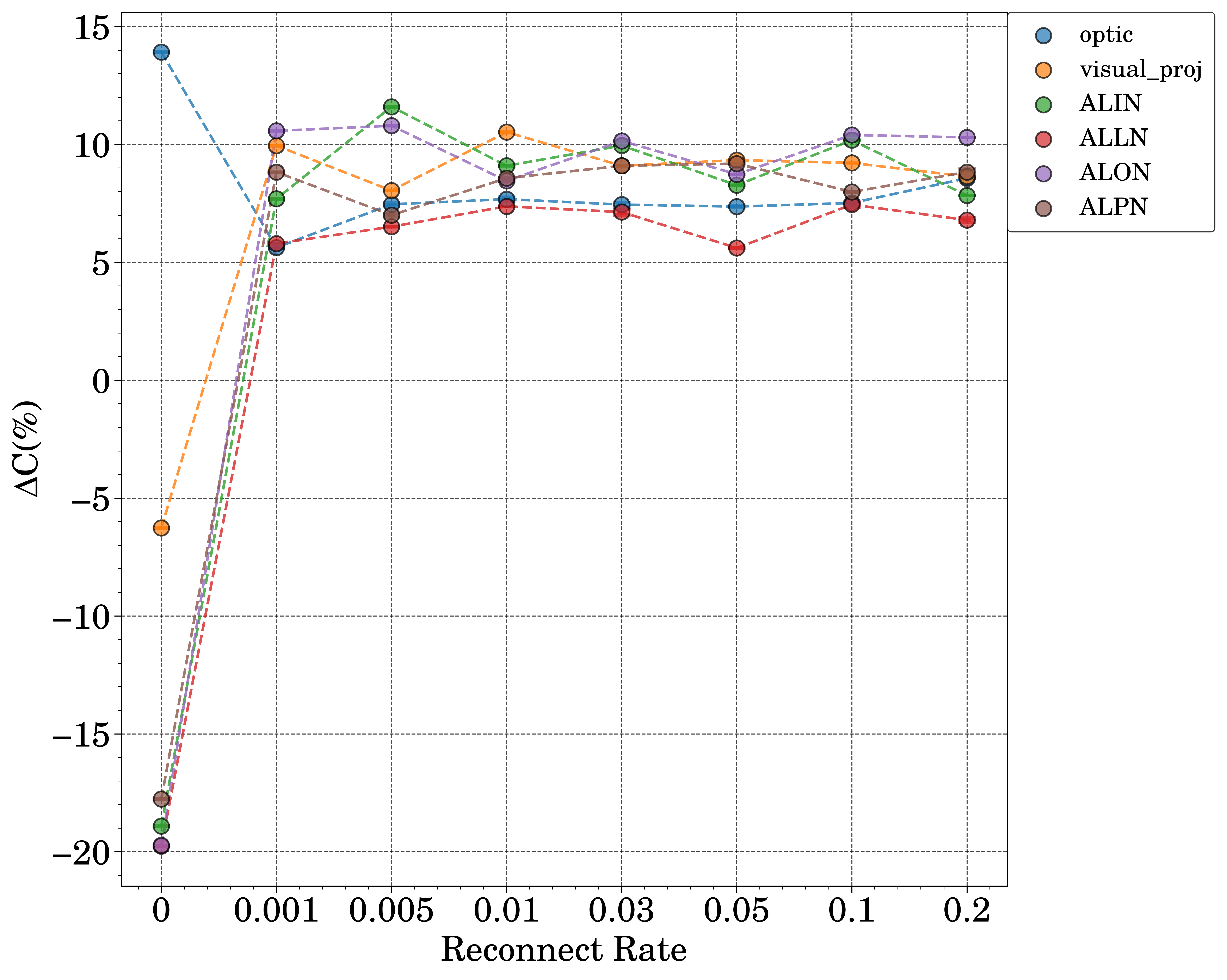}
		\end{minipage}		
		
	}
	\subfloat[sigmoid model, visual stimulus]{
		\begin{minipage}[t]{0.33\linewidth}
			\includegraphics[width=5.5cm,height=4.8cm]{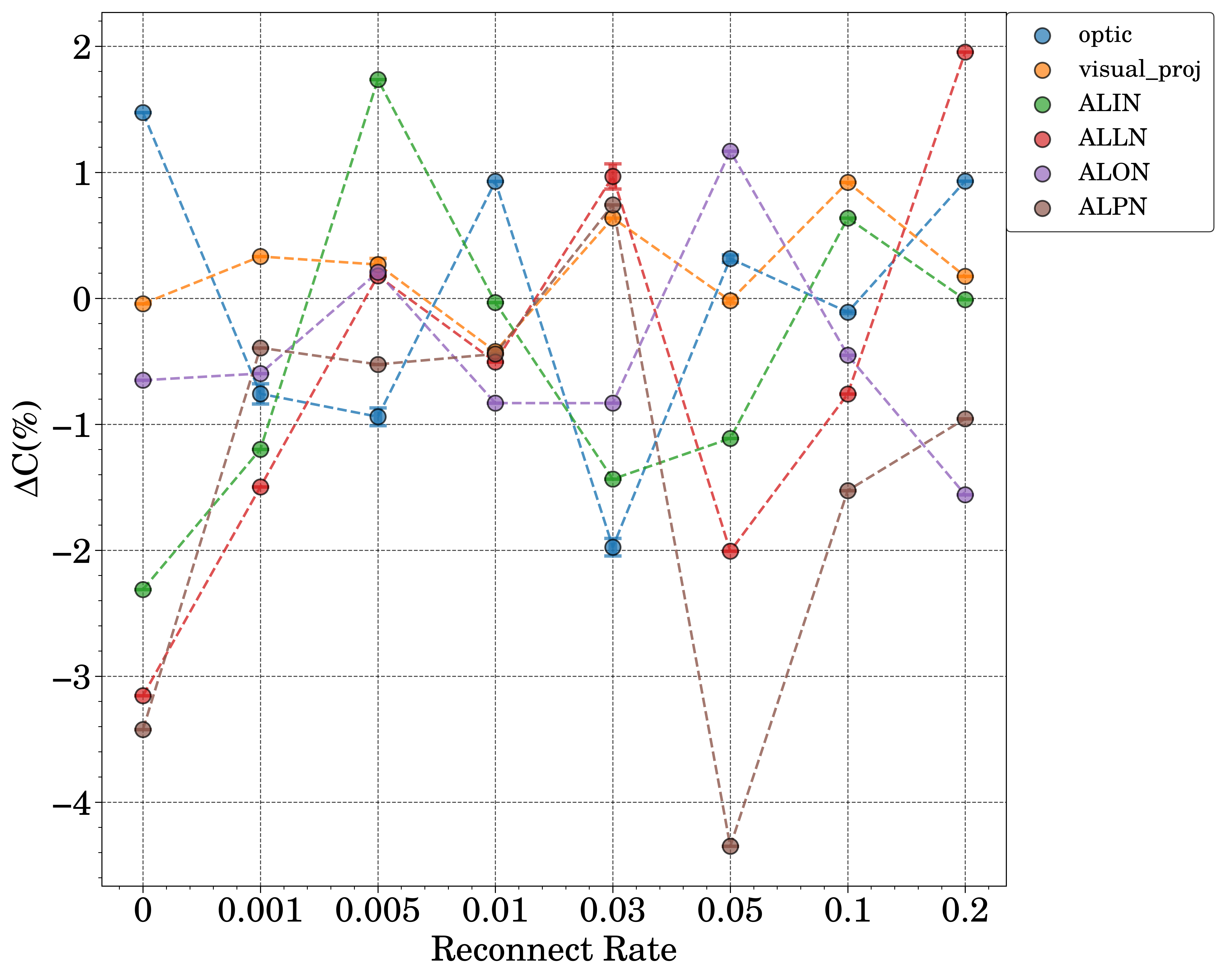}
		\end{minipage}		
	}
		\subfloat[LIF model, visual stimulus]{
		\begin{minipage}[t]{0.33\linewidth}
			\includegraphics[width=5.5cm,height=4.8cm]{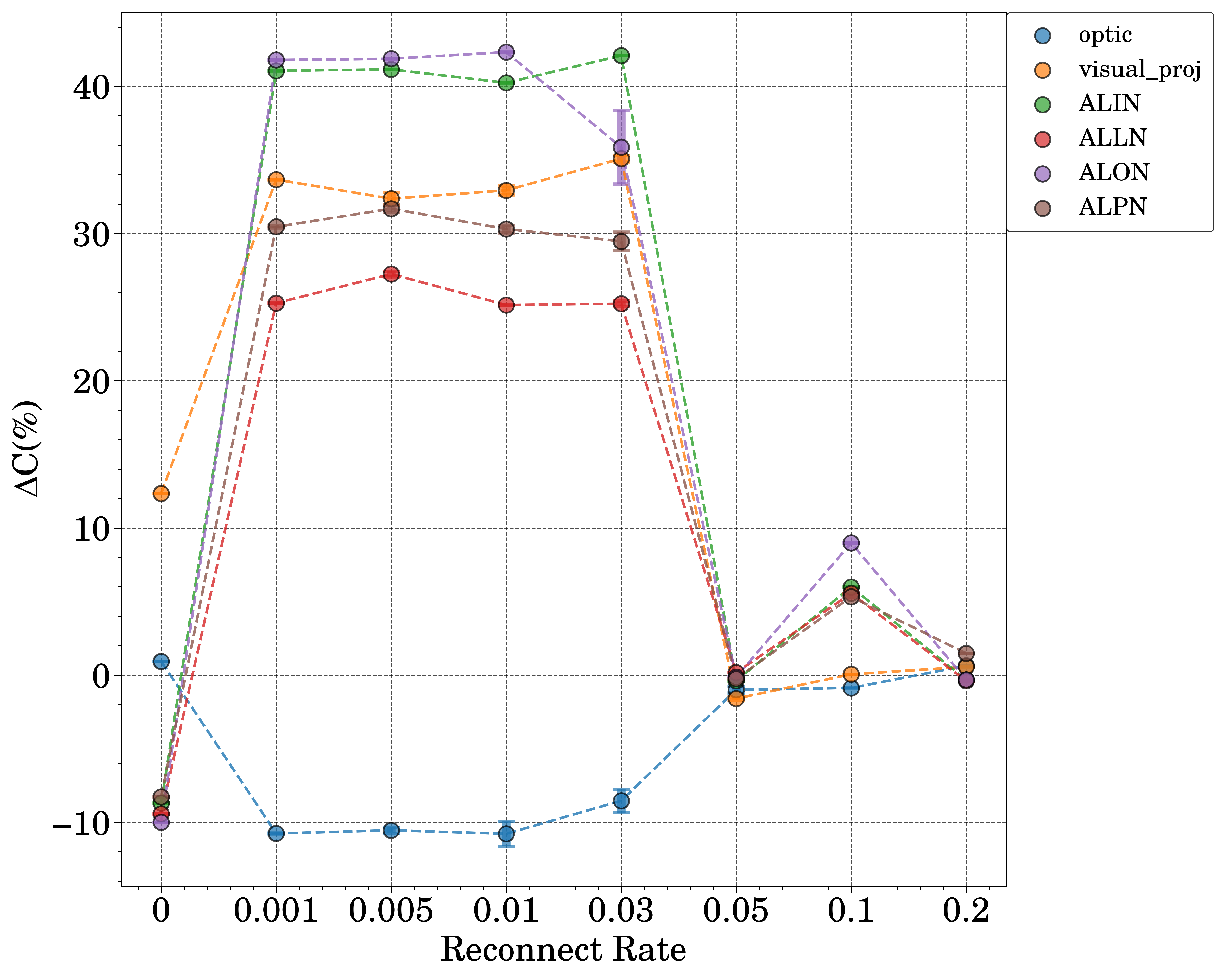}
		\end{minipage}		
	}
	\\
		\subfloat[threshold model, olfactory stimulus]{
		\begin{minipage}[t]{0.33\linewidth}
			\includegraphics[width=5.5cm,height=4.8cm]{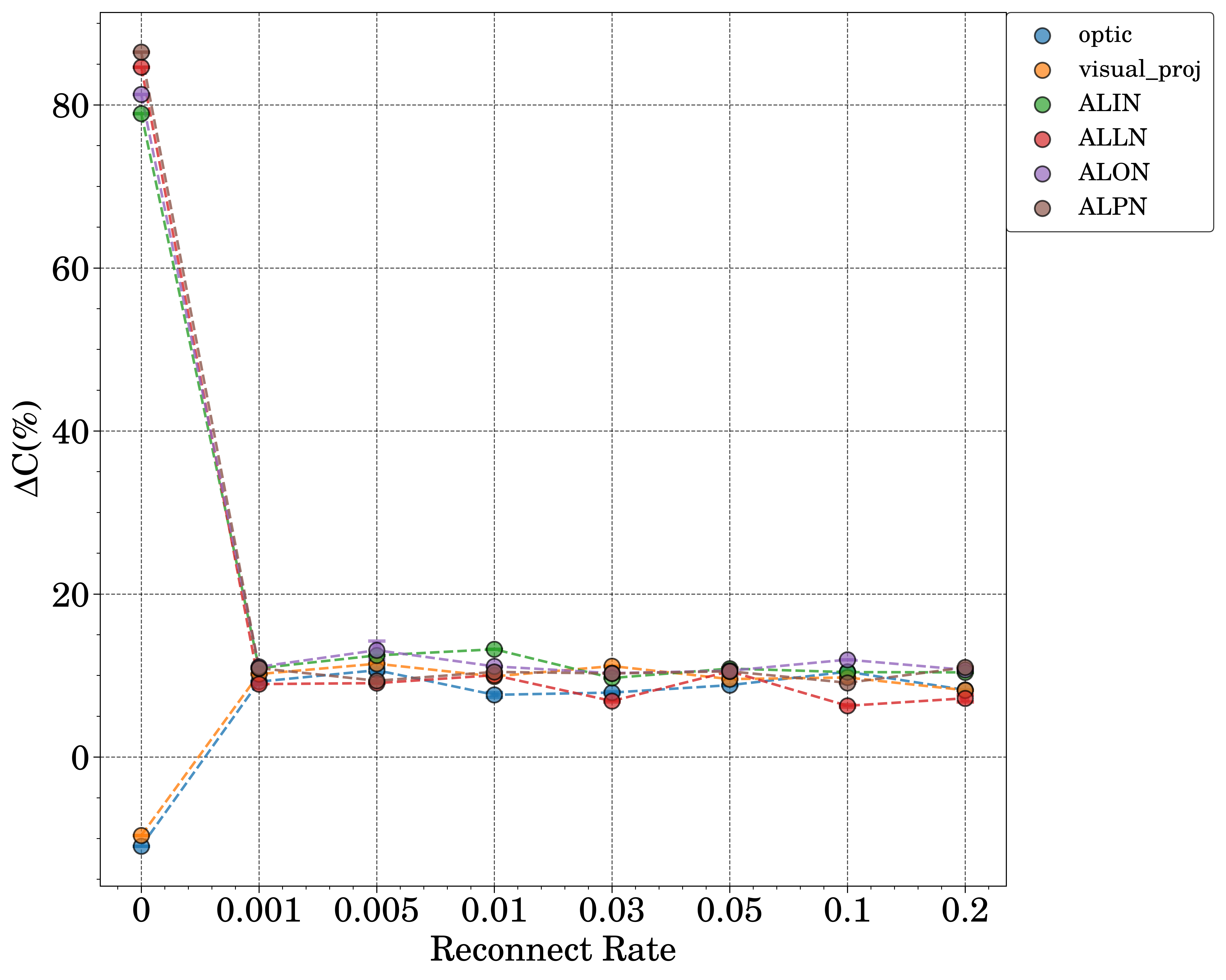}
		\end{minipage}		
		
	}
	\subfloat[sigmoid model, olfactory stimulus]{
		\begin{minipage}[t]{0.33\linewidth}
			\includegraphics[width=5.5cm,height=4.8cm]{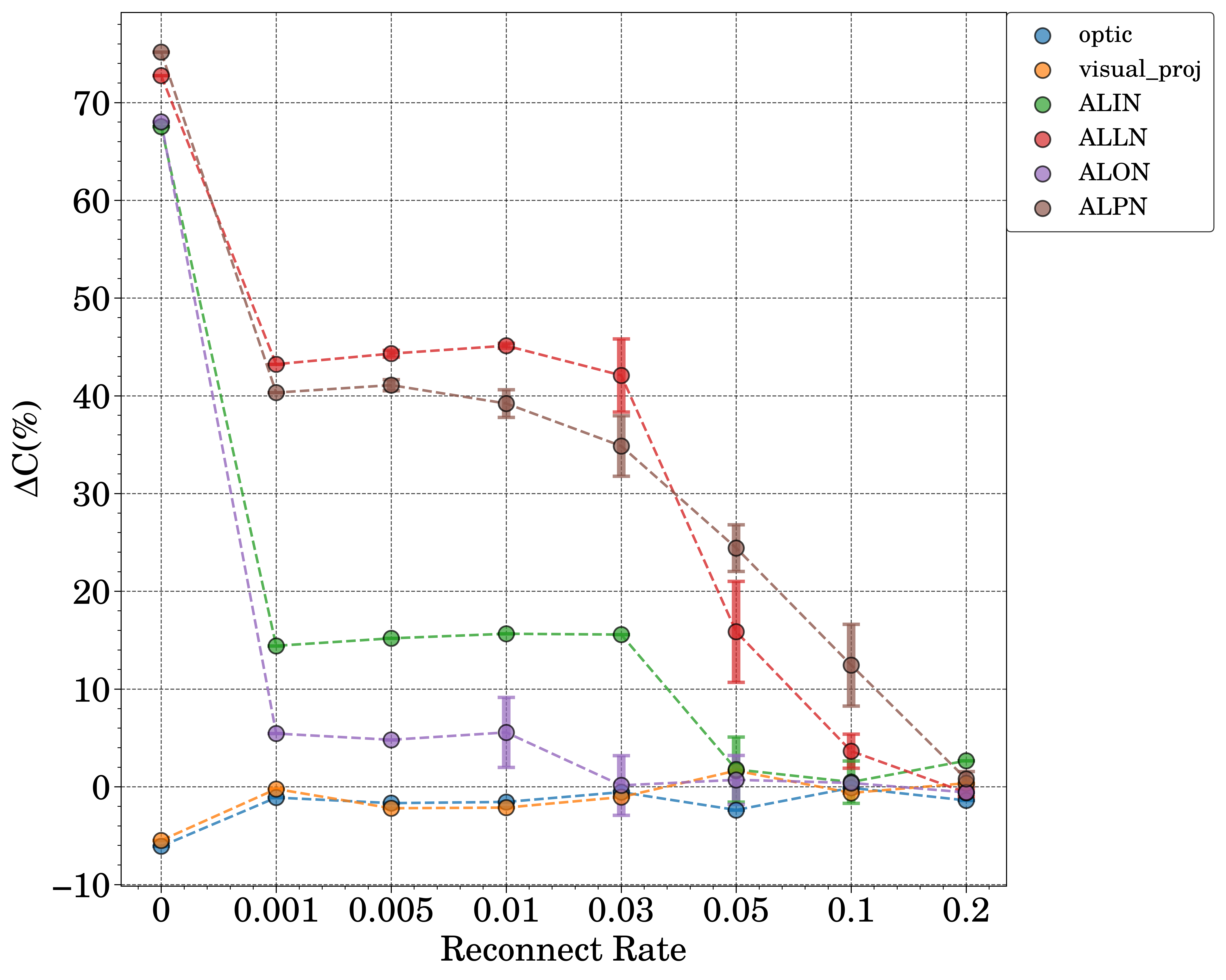}
		\end{minipage}		
	}
	\subfloat[LIF model, olfactory stimulus]{
		\begin{minipage}[t]{0.33\linewidth}
			\includegraphics[width=5.5cm,height=4.8cm]{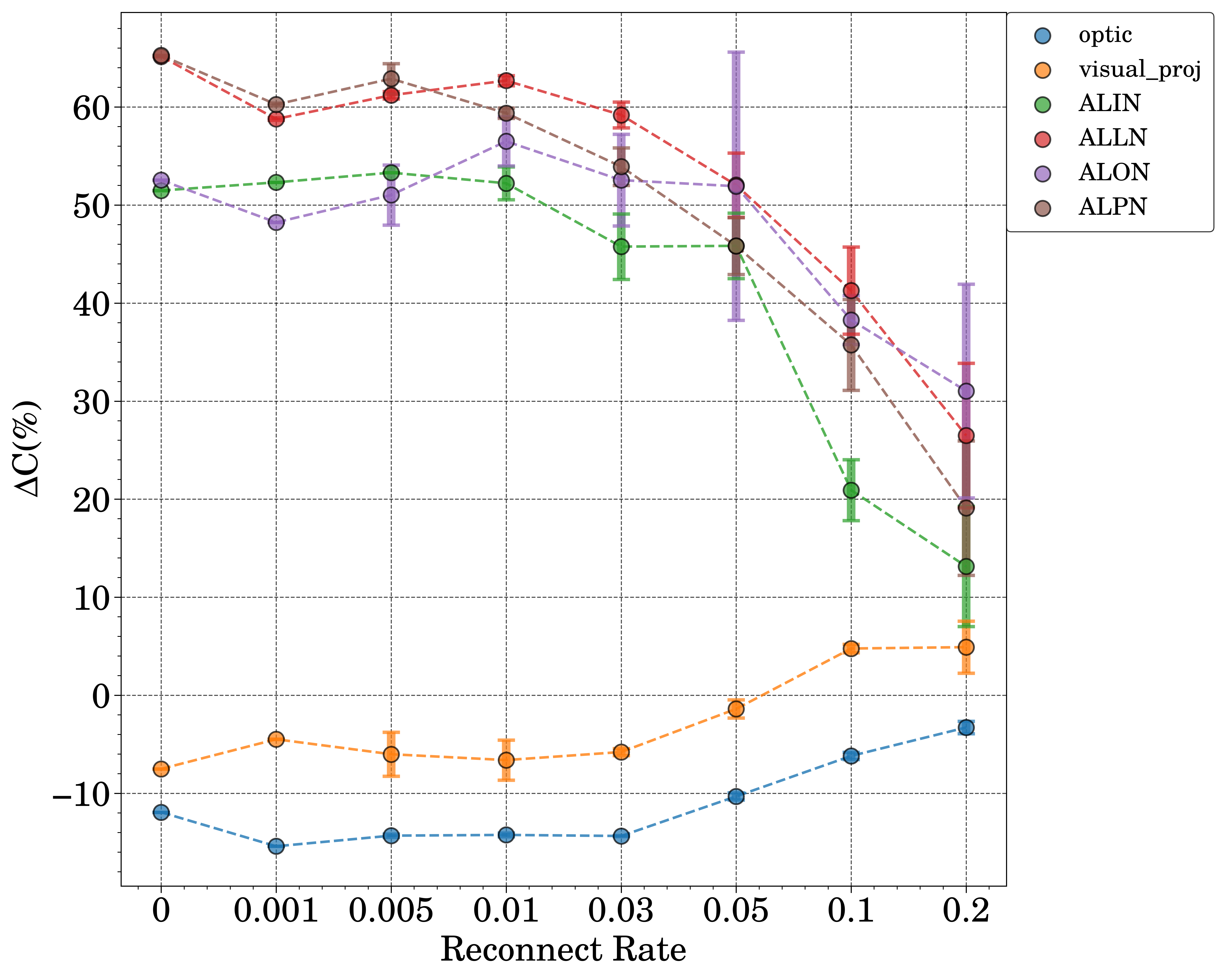}
		\end{minipage}		
	}
	\caption{Relationships of activation rate with reconnect rate. We conduct 5 repeated experiments and take the average and standard deviation.}
	\label{sensitivity}
\end{figure}

The experimental results are shown in Fig. \ref{sensitivity}, where 0 represents the original, unperturbed experimental results. At the 0 point, there is a significant difference in the relative activation ratios between visual and olfactory areas under specific stimuli. However, this significant activation pattern, discussed in the main text, disappears with a 1\textperthousand's network perturbation, replaced by a roughly equal activation ratio across all regions. This indicates that even a 1\textperthousand's change in network structure can lead to a rapid loss of network functionality specificity. Additionally, small-scale network perturbations (1\textperthousand-5\%) have a lesser impact on olfactory stimuli compared to visual stimuli. This is because the number of neurons associated with olfactory functions is significantly fewer than those related to visual functions, making it less likely for small-scale network structure changes to affect these neurons, thereby leading to a smaller impact on functionality.

Horizontal comparisons indicate that while there are no significant differences among different models in forming activation patterns, more complex neuron models better maintain their original activation patterns under network disturbances. This is because signals generated by complex neuron models have greater redundancy and functional diversity. For instance, the LIF (Leaky Integrate-and-Fire) model has a voltage decay mechanism, where the voltage gradually decays in the absence of a signal. Even if some input signals are erroneous, as long as the proportion of errors is not very large, the decay mechanism can filter out the impact of these erroneous signals.


\section{Large Spatial Network Visualization}\label{Visualization}

The visualization software of experiment observation on large spatial networks is important for researchers. However, there is still a lack of visualization system that can observe information propagation on large spatial networks in real-time, which poses difficulties in experiment. To observe directly the communication progress on this super large Drosophila connectome, we develop a large spatial network visualization software. This software will be made available to other researchers with the author's permission.

The drawing part mainly employs the Canvas provided by HTML5 and we use the Three.js based on WebGL technology to achieve spatial network visualization effects, while the user interface display is constructed by Vue.js, providing users with an intuitive and interactive interface to display and analyze data. 
\subsection{Front End Technical Details} 
We use the Vue.js framework to build the user interface. Vue.js is a popular JavaScript framework that simplifies the development of web applications through componentization, enabling the creation of front-end interfaces with rich interactivity and good user experience. 

The rendering of three-dimensional complex networks relies heavily on the Three.js library. Three.js is a powerful JavaScript library for creating and displaying 3D graphics. It offers a rich set of 3D objects, materials, lights, and animations, allowing for the easy creation of high-quality 3D scenes and objects.

To improve rendering performance, GPU acceleration technology is utilized. GPU acceleration takes advantage of the computational power of the Graphics Processing, significantly enhancing rendering speed and efficiency through parallel processing and stream processing modes. The GPU acceleration capabilities of Three.js, especially through the creation of point objects using Float32BufferAttribute, which contains their position information in three-dimensional space, greatly enhance the rendering speed and smoothness.

\subsection{Back End Technical Details} 
We use the Spring Boot framework, an open-source framework based on Java for quickly building independent, production-grade Spring applications. With Spring Boot, backend services can be rapidly set up to provide RESTful API interfaces for frontend consumption. The backend also utilizes the PostgreSQL database to store network data. 

Besides, We use PostgreSQL, which is a powerful open-source relational database management system that offers a rich set of data types and robust querying capabilities. To store 3D network data, we leverage PostgreSQL's JSONB type, which allows storing JSON-formatted data within PostgreSQL and provides a series of functions and operators for querying and manipulating this data. With the JSONB type, information about nodes and edges in the network can be stored, enabling fast querying and updating operations.
\subsection{Function Display}
\begin{figure*}[h]
	\centering
	\subfloat[Side view 1]{
		
		\includegraphics[width = 55mm]{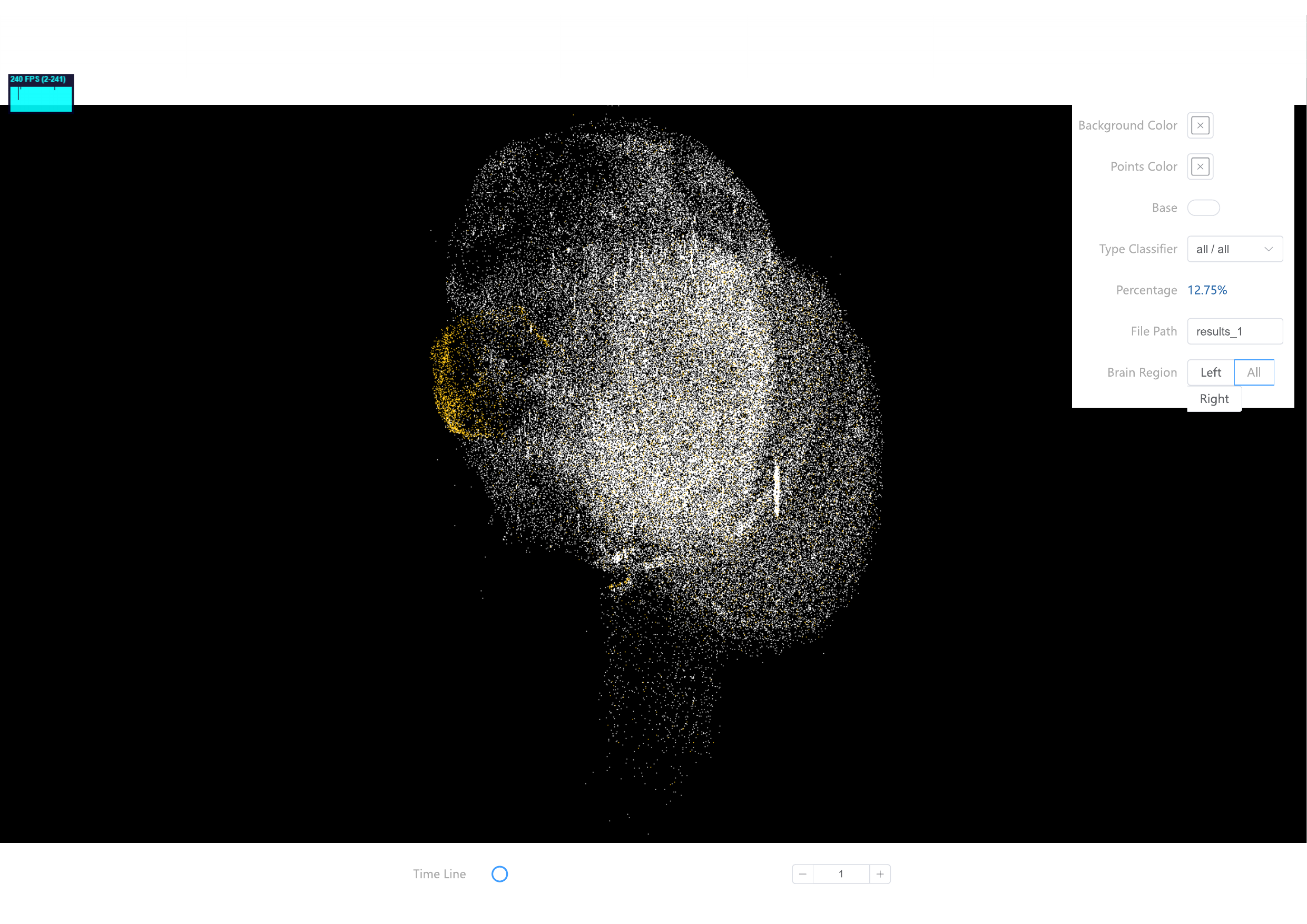}
		
	}
	\subfloat[Side view 2]{
		
		\includegraphics[width = 55mm]{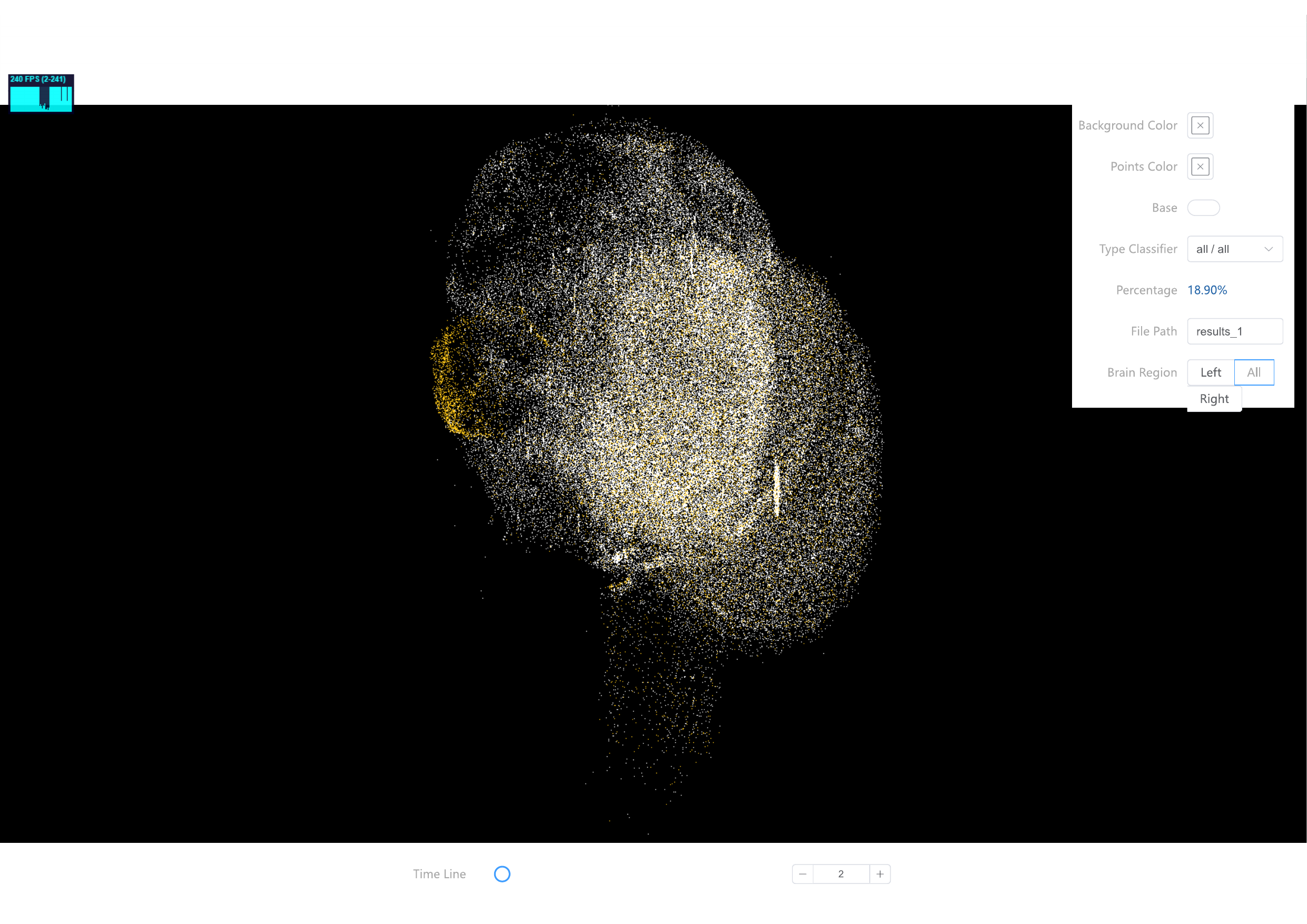}
		
	}
	\subfloat[Side view 3]{
		
		\includegraphics[width = 55mm]{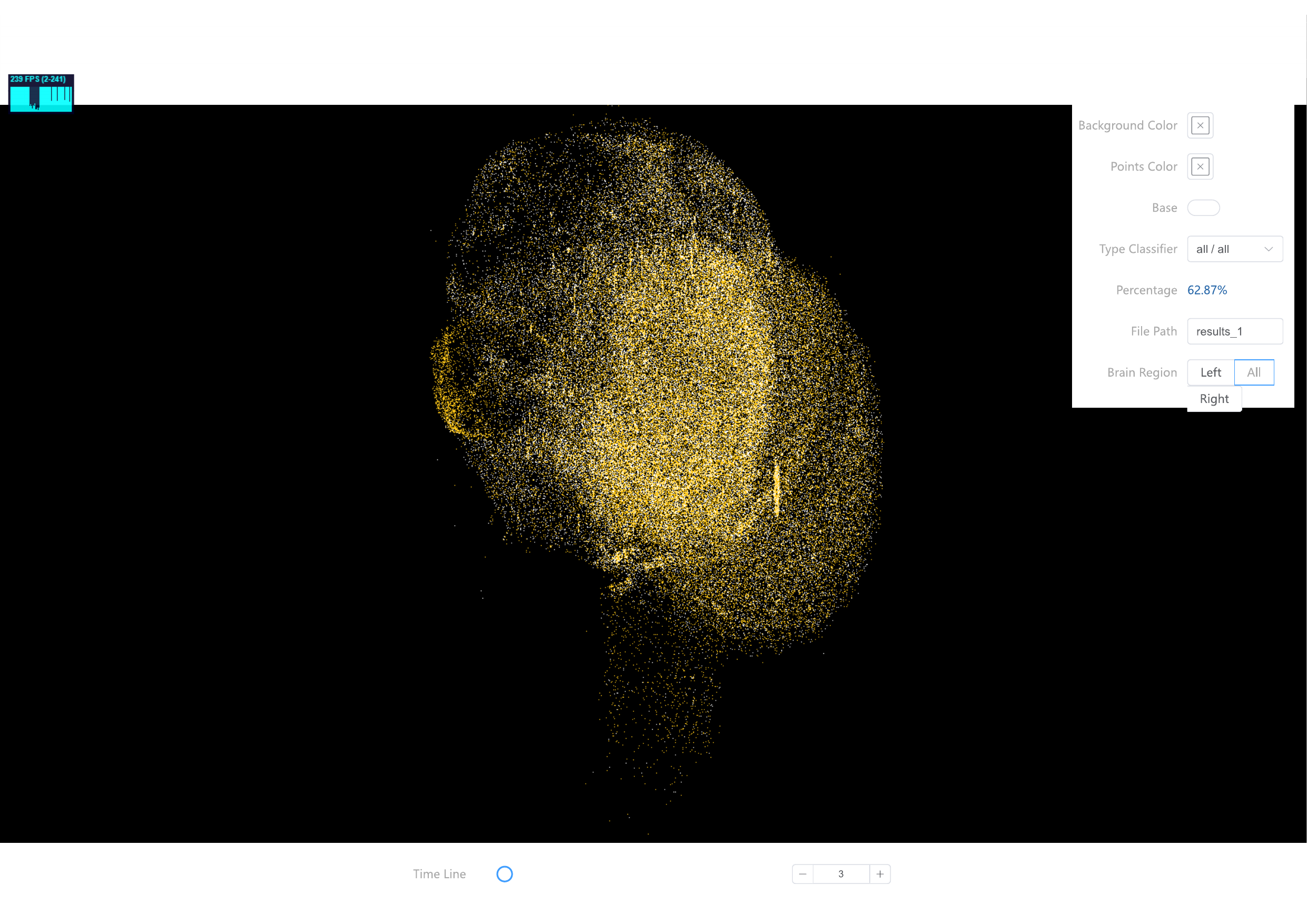}
		
	}
	\\
	
	\subfloat[Top view 1]{
		
		\includegraphics[width = 55mm]{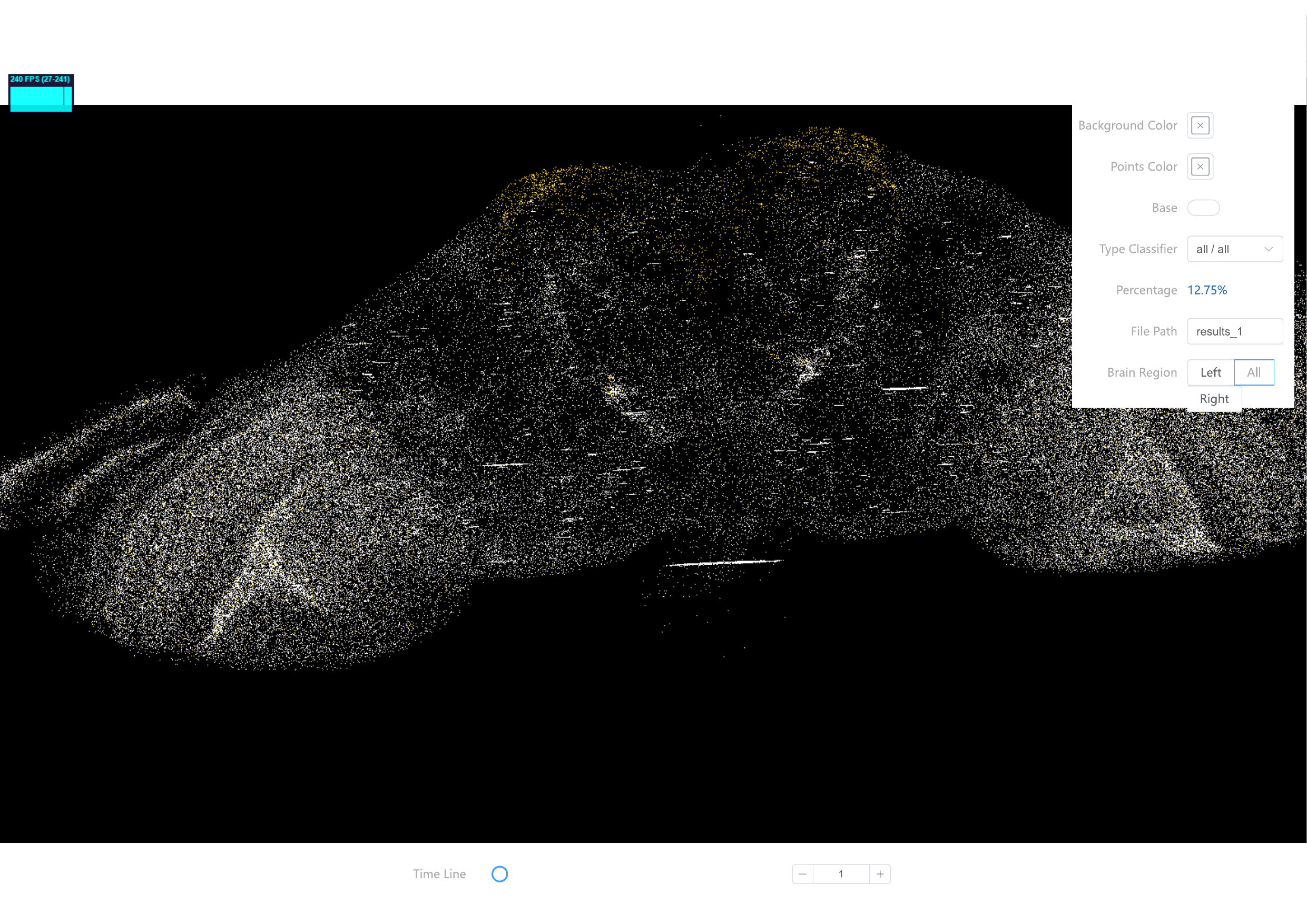}
		
	}
	\subfloat[Top view 2]{
		
		\includegraphics[width = 55mm]{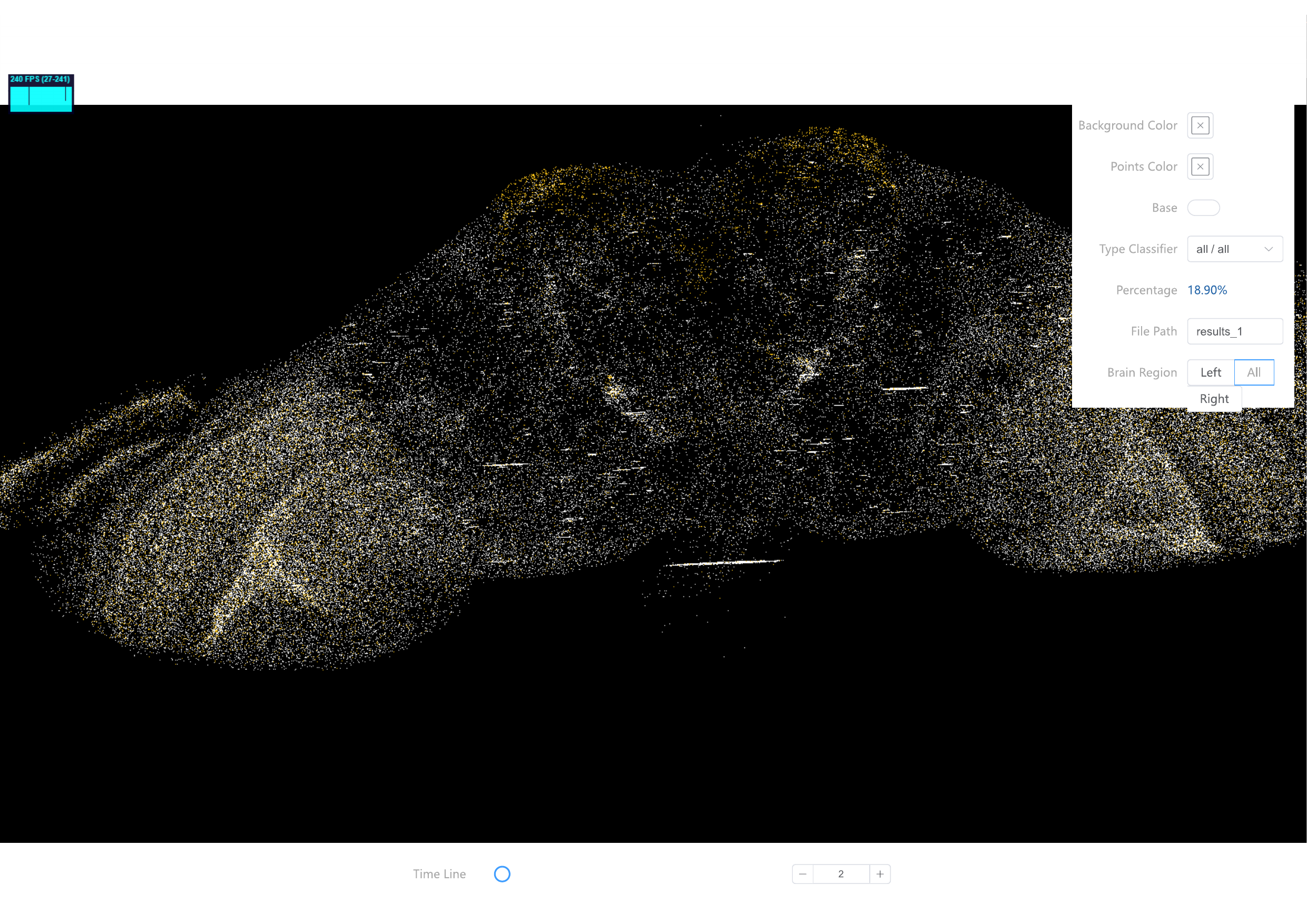}
		
	}
	\subfloat[Top view 3]{
		
		\includegraphics[width = 55mm]{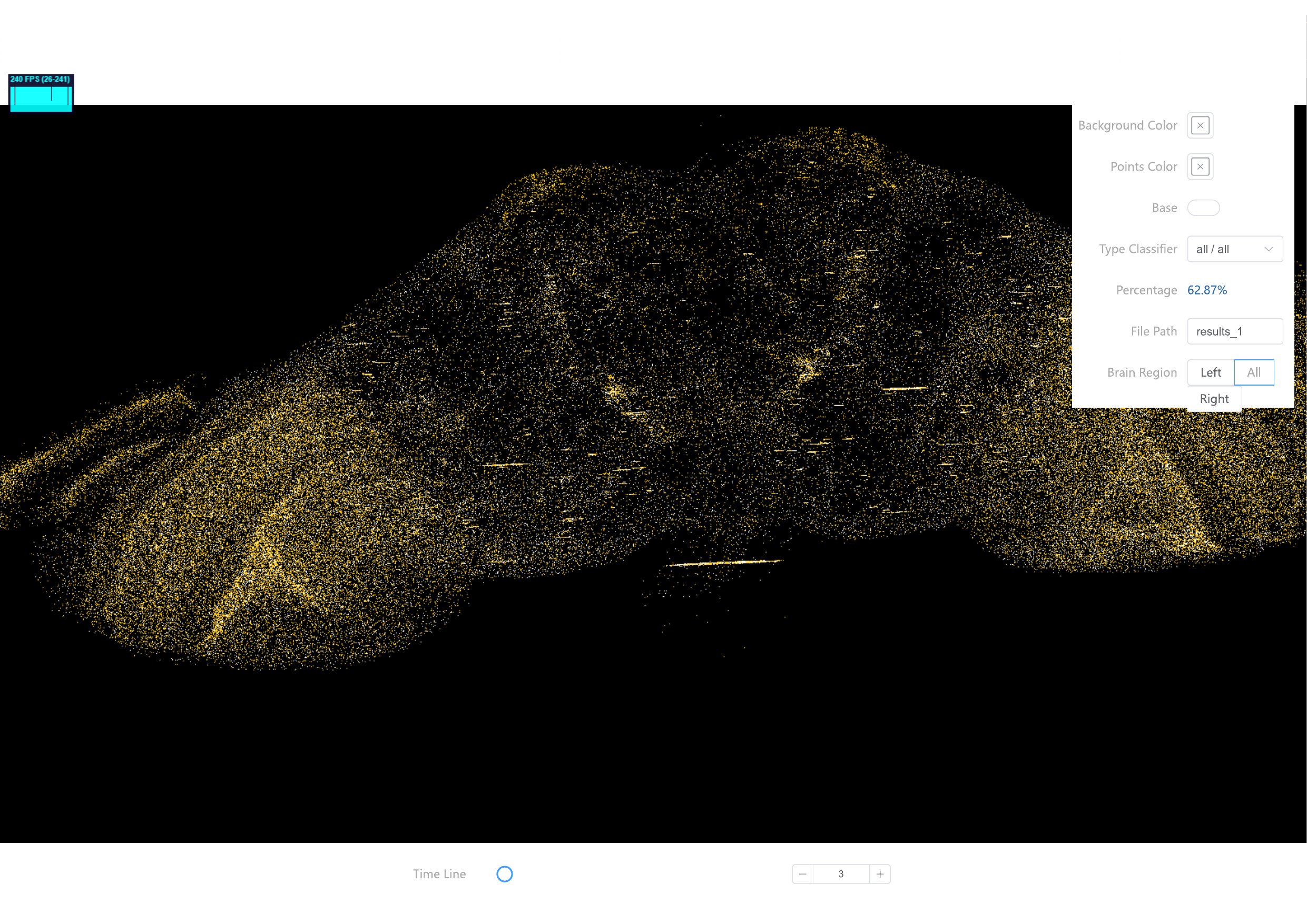}
		
	}
	\\
	\subfloat[Front view 1]{
		
		\includegraphics[width = 55mm]{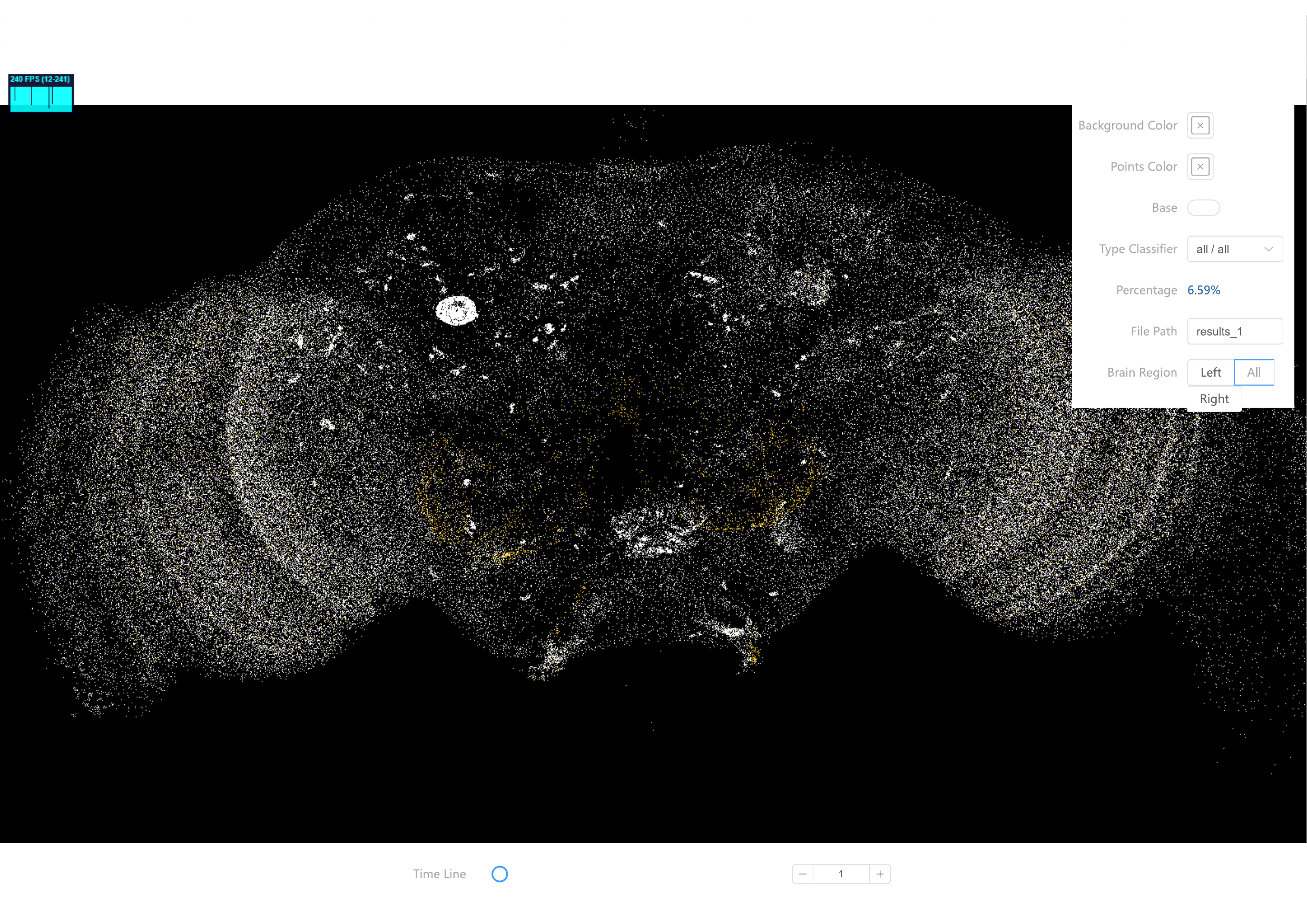}
		
	}
	\subfloat[Front view 2]{
		
		\includegraphics[width = 55mm]{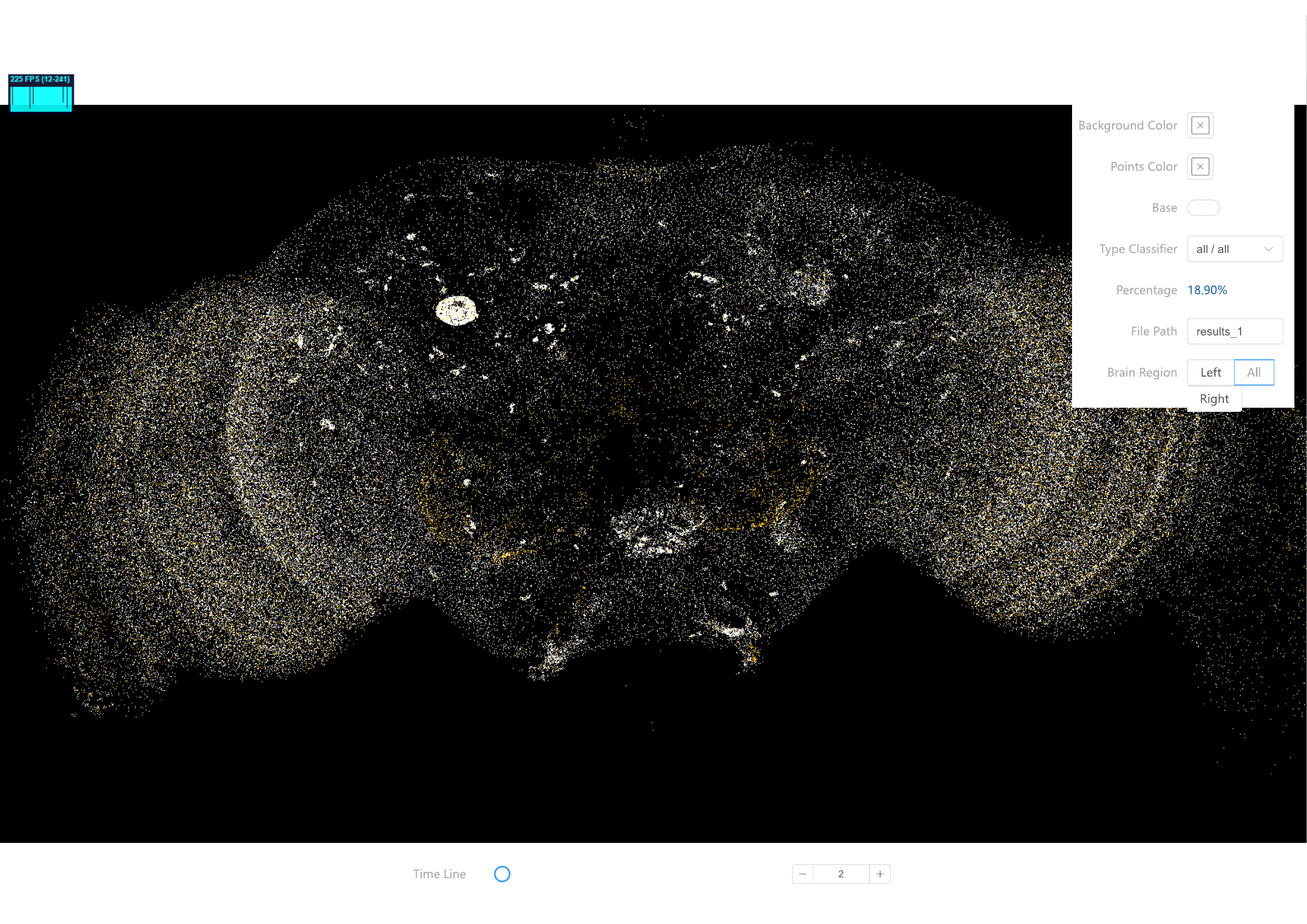}
		
	}
	\subfloat[Front view 3]{
		
		\includegraphics[width = 55mm]{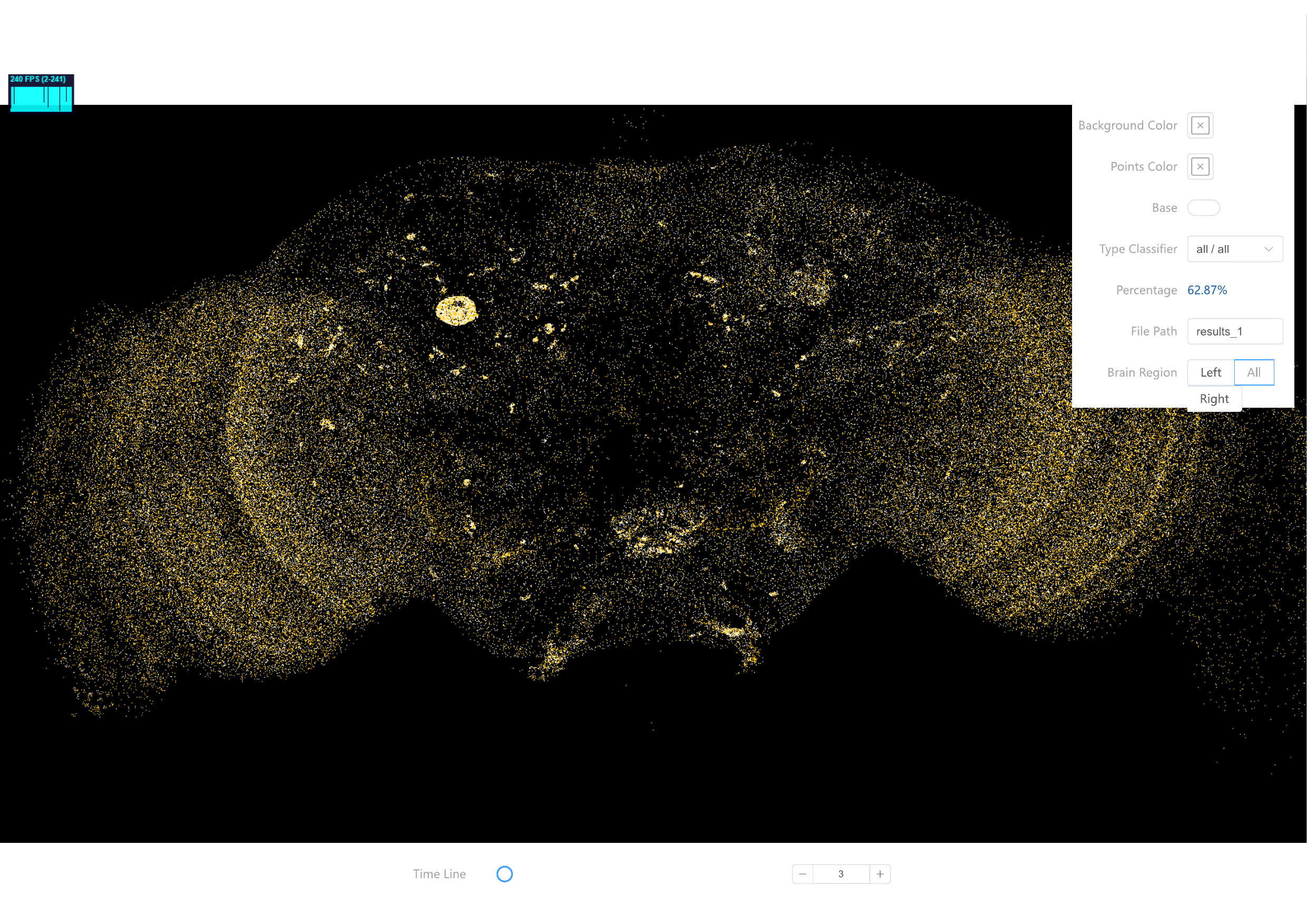}
		
	}
		%
		%
	\caption{Visualization of spatial Drosophila brain network.}
	\label{example}
\end{figure*}

\begin{figure*}[h]
	\centering
		\includegraphics[width = 150mm]{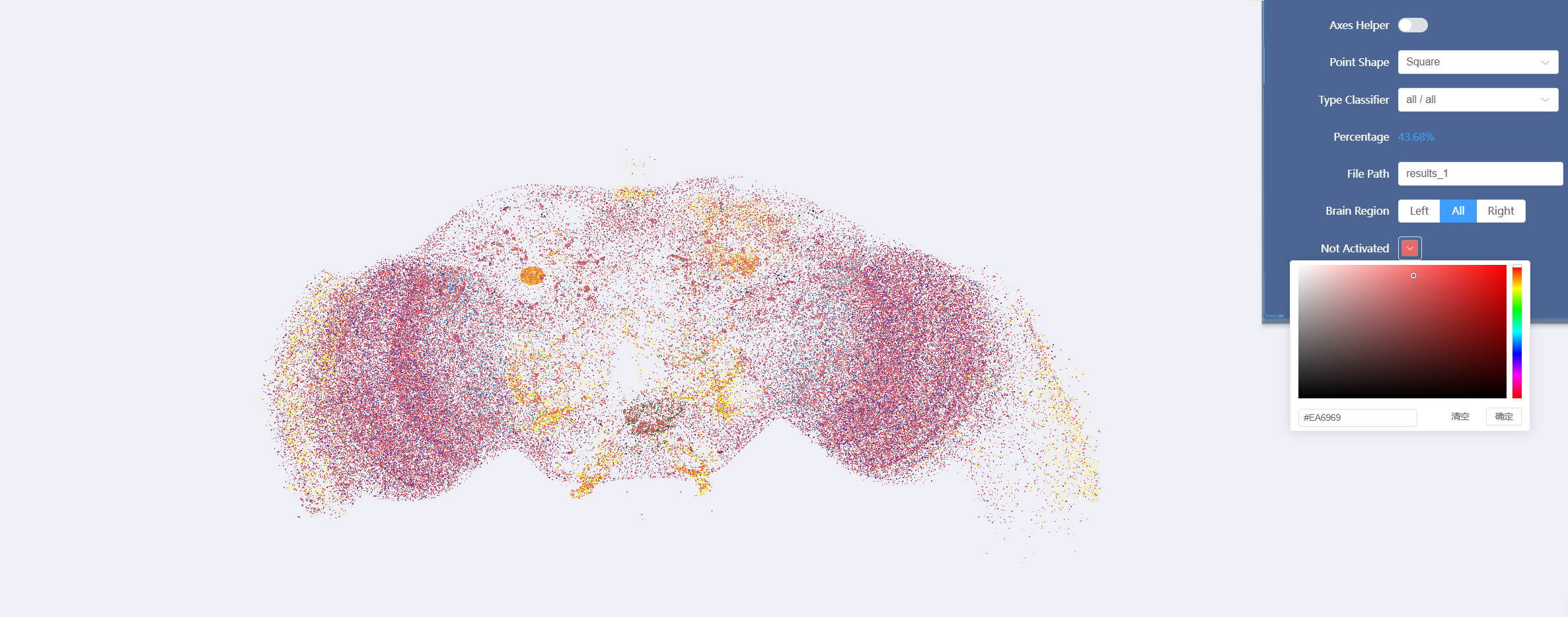}
	\caption{Display of color selection.}
	\label{display1}
\end{figure*}

Fig. \ref{example} is an example of the 3D visualization system. 
All nodes' colors can be adjusted by visible selection or input RGB code (Fig. \ref{display1}).

\begin{figure*}[h]
	\centering
	\includegraphics[width = 150mm]{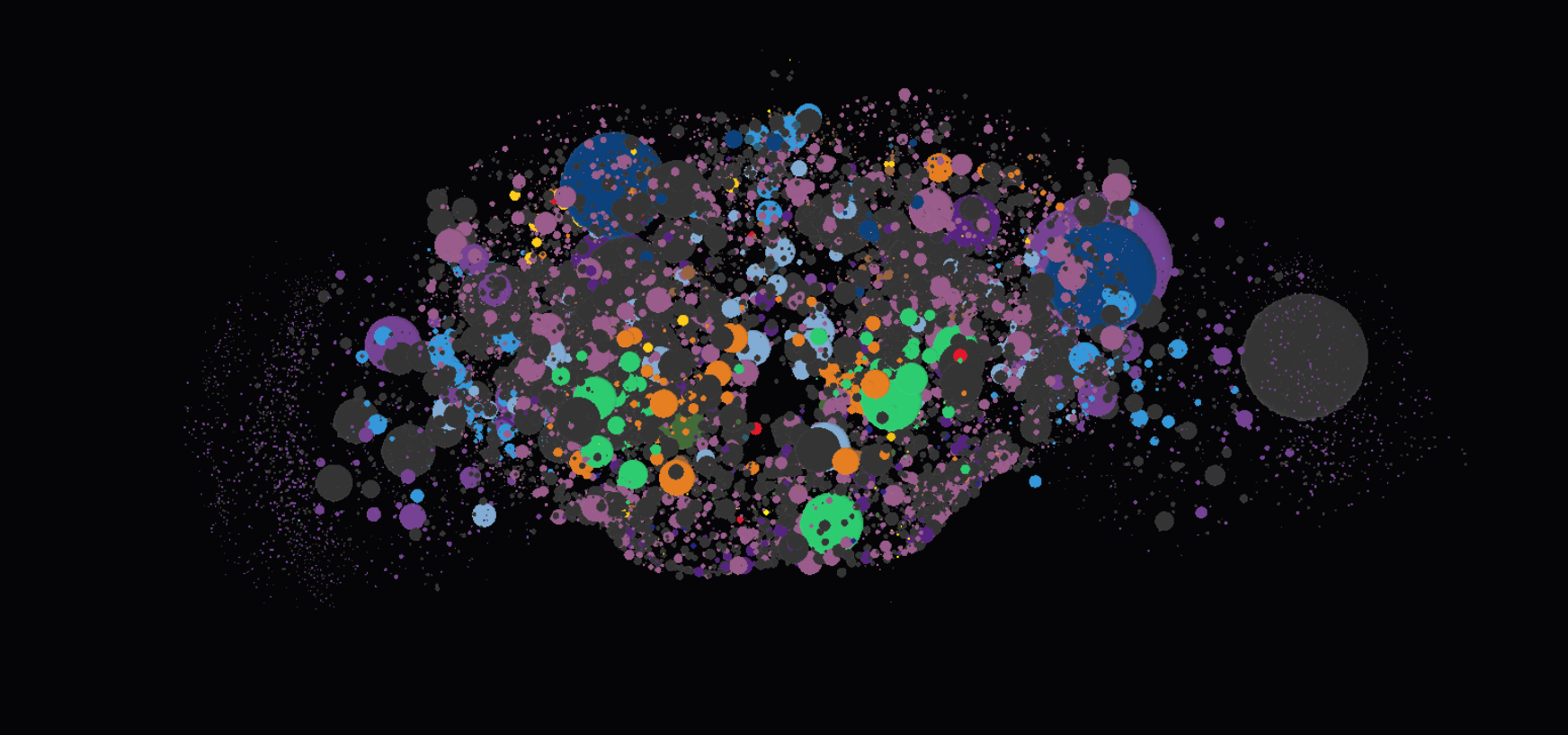}
	\caption{Display of nodes size adjusting.}
	\label{display2}
\end{figure*}

\begin{figure*}[h]
	\centering
	\includegraphics[width = 150mm]{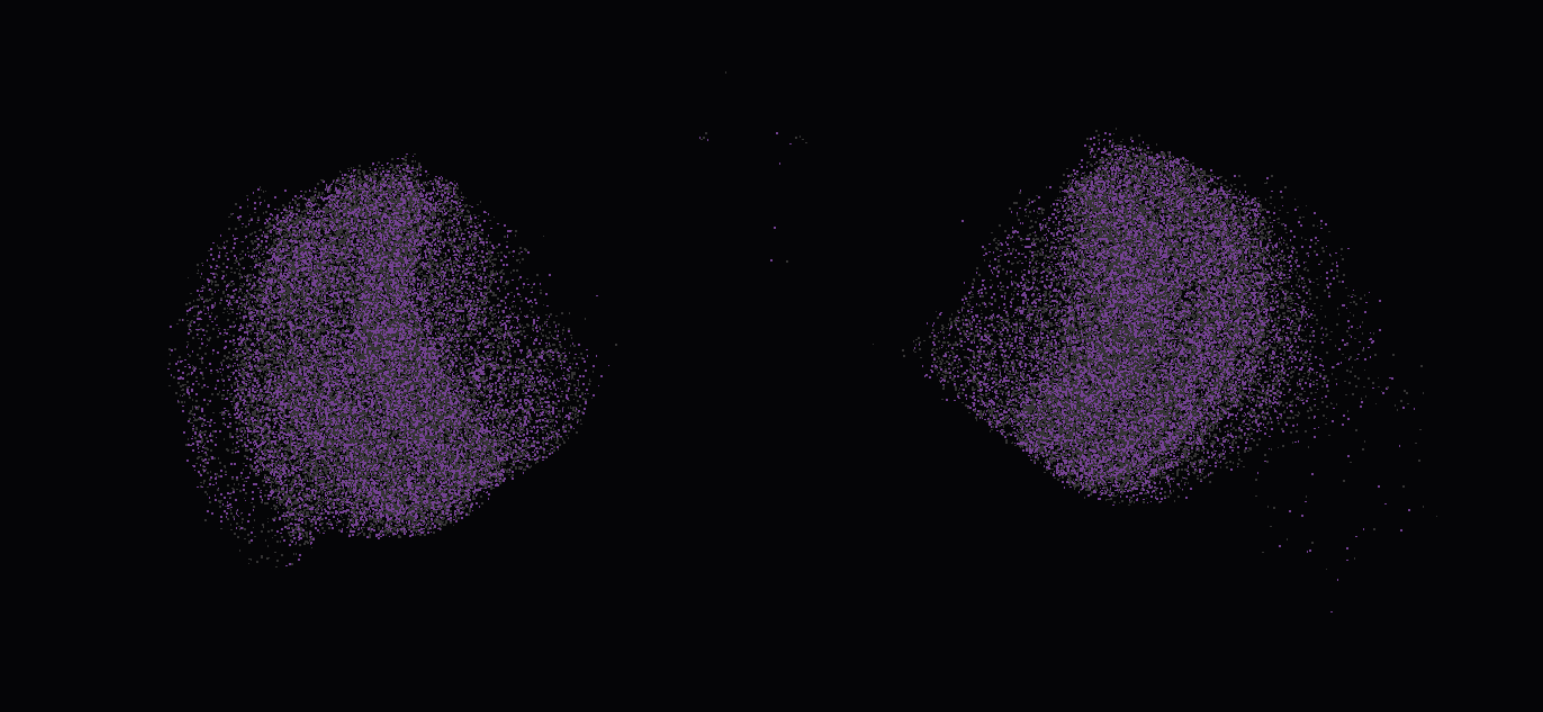}
	\caption{Display of area selection.}
	\label{display3}
\end{figure*}

Besides, from square to round, the shape of nodes can be adjusted even real neuron (Fig. \ref{display2}). 
It also can present different single areas (Fig. \ref{display3}).
\end{document}